\documentclass[aps,pra,twocolumn,superscriptaddress]{revtex4}
\usepackage[utf8]{inputenc}
\usepackage{times,graphicx,amsfonts,amsmath,amssymb,amsthm,mathtools,mathrsfs,xfrac,bbm,enumerate,booktabs,multirow,wrapfig,color}	
\usepackage[usenames,dvipsnames]{xcolor}

\renewcommand{\vec}{\mathbf}

\def\ep{{\cal E}}
\let\originalleft\left
\let\originalright\right
\renewcommand{\left}{\mathopen{}\mathclose\bgroup\originalleft}
\renewcommand{\right}{\aftergroup\egroup\originalright}
\renewcommand{\right}{\aftergroup\egroup\originalright}

\usepackage{cool}		%derivate integrali e non solo

\begin{document}

\title{Nonlinearity as a resource for nonclassicality in anharmonic
systems}
\author{Francesco Albarelli}
\affiliation{
Dipartimento di Fisica e Astronomia, Universit\`a di Bologna,
I-40127 Bologna, Italy}%\email{francesco.albarelli@studio.unibo.it}
\author{Alessandro Ferraro}%\email{a.ferraro@qub.ac.uk} 
\affiliation{School of Mathematics and Physics, Queen's University, Belfast
BT7 1NN, United Kingdom.}
\author{Mauro Paternostro}%\email{m.paternostro@qub.ac.uk} 
\affiliation{School of Mathematics and Physics, Queen's University, Belfast
BT7 1NN, United Kingdom.}
\author{Matteo G. A. Paris}%t\email{matteo.paris@fisica.unimi.it}
\affiliation{Dipartimento di Fisica, Universit\`a degli Studi di Milano, I-20133
Milano, Italy.}
\affiliation{CNISM, UdR Milano Statale, I-20133 Milano, Italy.}
\affiliation{INFN, Sezione di Milano, I-20133 Milano, Italy.}
\date{\today}
\begin{abstract}

Nonclassicality is a key ingredient for quantum enhanced technologies
and experiments involving macroscopic quantum coherence. 
Considering various exactly-solvable quantum-oscillator systems, 
we address the role played by the anharmonicity of their potential in the 
establishment of nonclassical features. Specifically, we show that a
monotonic relation exists between the the entropic nonlinearity of the 
considered potentials and their ground state nonclassicality, as quantified 
by the negativity of the Wigner function. In addition, in order to clarify the 
role of squeezing --- which is not
captured by the negativity of the Wigner function --- we focus on the Glauber-Sudarshan P-function and address
the nonclassicality/nonlinearity relation using the entanglement
potential. Finally, we consider the case of a generic sixth-order potential confirming the
idea that nonlinearity is a resource for the generation of nonclassicality and may serve as a guideline for the engineering of quantum oscillators.

%Nonclassicality is a key ingredient for quantum enhanced technologies
%and experiments involving macroscopic quantum coherence.
%We address the role played by inharmonicity of the system's potential in the 
%establishment of nonclassical features in the (ground) state of an oscillatory
%system. We show that a monotonic relation exists between the
%ground state nonclassicality, as quantified by the negativity
%of the Wigner function, and the entropic nonlinearity of relevant 
%potentials. Then, in order to clarify the role of squeezing, which is not
%captured by the negativity of the Wigner function, we look
%at singularity of the Glauber-Sudarshan P-function, and address
%the nonclassicality/nonlinearity relation using the entanglement
%potential. We address the case of a generic potential confirming the idea that nonlinearity is a resource for the generation of nonclassicality and may
%serve as a guideline for the engineering of quantum oscillators.
\end{abstract} 
\maketitle
%%%
\section{Introduction}
\label{sec:intro}
At the heart of quantum technologies lies the fact that quantum mechanical systems show features, with no classical counterpart, that may be employed as resources to perform specific tasks better or faster than within the classical realm~\cite{NC10}. 
%Coherence of quantum systems, together with
%its multipartite manifestation, entanglement, have indeed been 
%recognized the most important resources for quantum information
%processing.
%\par
In the context of quantum optics, genuine quantum traits of optical systems have led to the emergence of the concept of  \emph{nonclassicality}, which characterizes states whose effects are not achievable with classical light~\cite{MW95}. In particular, \textit{linear} models (here intended as systems that induce linear transformations of the bosonic mode operators) have attracted much attention in the past decades, due to the development of experimental platforms able to implement them. In fact, the generation of non-classical light, especially in the form of squeezed beams, has proven to be an enabling resource for a variety of  quantum technological applications \cite{L14}.

Recently, alternative experimental platforms have been developed that can also be coherently controlled and described as single-mode bosonic systems --- including trapped ions \cite{L03}, optomechanical systems \cite{AKM14}, atoms in optical lattices \cite{LSA13}, and hybrid systems \cite{R+14}. The latter naturally embody a playground to discuss and test the generation and  characterization of genuine quantum features. In particular, they offer the unique opportunity to consider \textit{nonlinear} (or anharmonic) models, given that the possibility to host non-linearities is within reach of current technologies, in particular for trapped ions \cite{H+11} and optomechanical systems \cite{S+09}. Interestingly, it has been shown that the inclusion of nonlinearities in the oscillator potential uncovers new possibilities to generate nonclassical states \cite{DIVINCENZO, ONG, PEANO, KOLKIRAN, ANDERSSON,RIPS,Vacanti2013,MFB14}. However, a general framework that encompasses these possibilities remains elusive and in particular a thorough quantitative assessment of the link between nonlinearity and nonclassicality still lacks.

The aim of this work is to investigate in details the idea that
nonlinearity is a general resource to generate nonclassicality in
single-mode bosonic systems constituted of anharmonic oscillators. In
particular, we will focus on a quantitative assessment of the
phenomenon, as we critically consider specific quantifiers of
nonclassicality and nonlinearity. In fact, identifying proper measures
of these quantities is crucial by itself and, in particular, different
figures of merit exist that  capture different features associated to
nonclassicality
\cite{H87,Lee1991,RV02,Kenfack2004,Asboth2005,Mari2011,FP12}. The
quantitative connection of the nonlinear behaviour of an oscillatory
system and the appearance of nonclassicality has recently been tested,
in the context of nano-mechanical resonators, for the Duffing oscillator
model \cite{Teklu2015}. Here we extend such connection and assess its
validity for more general scenarios, including three families of exactly
solvable non-linear oscillators and a generic sixth order potential.
\par
The remainder of this paper is structured as follows.  In Sec.~\ref{sec:nonclassnonlin}
we review the main conceptual tools and establish our notation and formalism. First, we
introduce and discuss the two quantitative measures of nonclassicality
that will be used throughout the paper, namely entanglement potential
and the volume of the negative part of the Wigner function. Then, we
review a recently introduced measure to quantify the nonlinearity of a
quantum oscillator \cite{Paris2014}, which in turn is based on an
entropic measure of non-Gaussianity~\cite{Genoni2010}.  In Section
\ref{sec:potentials} we analyse the quantitative connection between
nonclassicality and nonlinearity for three different nonlinear
potentials having an exact solution. We also highlight
some differences between the two measures of nonclassicality (see also 
Ref.~\cite{Li2010}). In Sec.~\ref{sec:harmoscpert} we address 
generic (symmmetric) anharmonic potential by considering 
fourth- and sixth-order perturbations to the harmonic one. 
In Sec.~\ref{sec:out} we draw our conclusions.

\section{Nonclassicality of a state and Nonlinearity of a potential}
\label{sec:nonclassnonlin}

\subsection{Nonclassicality of a Single-mode Bosonic State}
We consider a bosonic system with a single degree of freedom, 
such as a one-dimensional oscillator or a single-mode of 
a bosonic field. Since we deal with single-mode systems, we will not discuss 
any notion related to entanglement or other nonclassical correlations.

In the most general terms, a quantum state is said to be nonclassical if
the methods of classical statistics fail to describe its properties and
phenomenology.  In the context of quantum optics this definition is made
precise by using quasiprobability distributions in phase space.
Here we are not only interested in \emph{criteria} for
nonclassicality, but we seek for a \emph{quantitative characterization}.  
An excellent summary on this topic can
be found in the introduction of Ref. \cite{Miranowicz2015}.

\subsubsection{P-nonclassicality and Entanglement Potential}
\label{subsec:ep}
According to Titulaer and Glauber~\cite{Glauber1963,Titulaer1965,Mandel1986}, a quantum state is 
nonclassical when its $P$ function fails to be interpreted as a
probability distribution in the phase space (see also Refs.~\cite{Vogel2006,Johansen2004,Spekkens2007}).
It has been recently emphasized~\cite{Kiesel2013} that the $P$ function
is the only quasiprobability distribution which can give a description
that can be completely modelled using classical electrodynamics,
therefore supporting the idea that to identify a classical state it is
necessary to use the $P$ function. In this paper we dub such
fundamental notion as {\it P-nonclassicality}. 

The best known way to quantify P-nonclassicality is the nonclassical
depth~\cite{Lee1991}: It quantifies, operationally, the amount of thermal noise 
that is needed in order to render the $P$ function of a given state a well-behaved probability
distribution and the corresponding state classical.  This measure
however is not fully suited for our purposes: in fact, while we will be interested in establishing a quantitative hierarchy 
of pure non-Gaussian states in terms of their nonclassicality at a set nonlinearity of a given potential, it has been proven
that such states all saturate the nonclassical depth~\cite{Lutkenhaus1995},
i.e. they are equally and maximally nonclassical according to this measure.

This obstacle can be overcome by considering the following. It has long been 
known that coherent states are the only pure states
that produce uncorrelated outputs when mixed by a passive linear-optics
device~\cite{Aharonov1966}. Specifically, P-nonclassicality has been
identified as a necessary condition for having entangled states at the output of 
a beam splitter~\cite{Kim2002,Xiang-bin2002} and quantitative relations have 
been identified between non-classicality and entanglement 
\cite{WEP:03,Asboth2005,oli09,oli11,JLC:13,VS:14,B+15} or discord-like 
correlations more in general \cite{B+15}. The idea of quantifying 
nonclassicality of a single mode state as the
two mode entanglement at the output of a linear optic device was
introduced by Asb\'oth et al.~\cite{Asboth2005}. In particular, it was shown 
that the optimal entangler is just a (50:50) 
beam splitter with vacuum as an auxiliary state. By restricting to this setup,
nonclassicality of the input state becomes a necessary and
sufficient condition for output entanglement. As a consequence,
entanglement at the output of a beam splitter may be used as a
faithful quantitative measure of P-nonclassicality. This measure  
is usually referred to as entanglement potential ${\cal E}(\rho)$ and it is defined as 
\begin{equation}
\label{eq:EntPotDef}
\ep[\rho]=E\left[ \hat{B} \left( \rho  \otimes |0 \rangle \langle 0 | \right) \hat{B}^{\dag} \right],
\end{equation}
where $\rho$ is the density matrix of the state under scrutiny,
$|0\rangle$ is the vacuum state at the ancillary port of the beam
splitter, $\hat B$ is the beam splitter operator, and $E[\rho]$ is a
suitable measure of entanglement. 
Our analysis will be concerned with the ground state of a given
Hamiltonian model. By dealing with pure states, $E[\rho]$ can be chosen,
with no ambiguity, as the the entanglement entropy. This choice
corresponds to the \emph{entropic entanglement potential} defined in
Ref.~\cite{Asboth2005}, which has been evaluated by truncating the
dimension of the Hilbert space to a suitable dimensione, ensuring
the normalization of the state before and after the beam-splitter.
%
%We truncate this series to a finite number $N$ and as a validity check
%of the approximation we have to make sure that the norm (squared) of
%the vector \eqref{eq:MHOseriesvec}, that is to say the sum
%$\sum_{n=0}^{N} |\phi_n |^2$, is sufficently close to unity.  
%The actual computation of the entropy relies on
%a numerical diagonalization of the reduced density matrix written in the
%Fock basis.

\subsubsection{$W$-nonclassicality}
\label{subsec:negwig}
While the $P$ function can be a singular object, 
the Wigner function is always well behaved, even if it can attain negative values.
Negativities of the Wigner function associated with a given state defines the so-called {\it $W$-nonclassicality}, 
which is, however, only a sufficient condition for P-nonclassicality.
It follows that there are W-classical states which are P-nonclassical:
displaced squeezed states are a remarkable example. The notion of $W$-nonclassicality has gained an operational
meaning as follows: the evolution of a system which is in a $W$-nonclassical state 
cannot be efficiently simulated with classical resources~\cite{Mari2012,Veitch2013}. In order to quantify $W$-nonclassicality we use the volume 
of the negative part of the Wigner function~\cite{Kenfack2004}
\begin{equation}
\label{eq:def_delta}
\delta = %\int \! \mathrm{d}^2 \alpha \, \left[ |W(\alpha)| - W(\alpha) \right] = 
\left( \int \! \mathrm{d}x\,dp \, |W(x,p)| \right)  - 1,
\end{equation}
where $x$ and $p$ are phase-space variables, and $W(x,p)$ is the Wigner function of the state under scrutiny. We will make use of the following normalized version of this measure
\begin{equation}
\label{eq:def_nu}
\nu = \frac{\delta}{1+\delta},
\end{equation}
which gives $\nu\in[0,1]$.

Let us stress that the $W$ and $P$-nonclassicality single out different quantum features. In particular, the Hudson theorem~\cite{Hudson1974} guarantees that the sole pure states with a positive Wigner function are Gaussian ones, i.e. squeezed coherent states. Hence, there exist pure states that have zero $W$-nonclassicality (\textit{e.g.}, squeezed states) but non-zero $P$-nonclassicality. In this sense the entanglement potential can reveal more detailed features of quantumness, as we will see below. Note that measures of $W$-nonclassicality based on the geometric distance between quantum states have also been introduced \cite{H87,Mari2011}.

\subsection{Quantifying the Nonlinearity of a One-dimensional Potential Using Its Ground State}
\label{subsec:nonlin}
The first idea to quantify the nonlinearity (inteded as the
\emph{anharmonicity character}) of a potential would be defining a
distance between potential functions and the reference harmonic
potential. However, this is in general not feasible, since
potentials do not need to be integrable functions. A different approach
follows from the fact that ground states and equilibrium states 
of anharmonic potentials are not Gaussian,
as opposed to those of a quantum harmonic oscillator. We can
thus choose to quantify nonlinearity by the non-Gaussianity of the ground
state of a given Hamiltonian model~\cite{Paris2014}. The measure of non-Gaussianity
used for this goal is the entropic measure introduced in
\cite{Genoni2008,Genoni2010}. Here we shall briefly review these
measures.

\subsubsection{Non-Gaussianity of a Quantum State}
%Gaussian states are $n$ modes bosonic states with a Gaussian Wigner function \cite{Ferraro2005}, but we will deal only with the simplest case of a single mode state, so we just have a single pair of creation and destruction operators $\hat{a}$ and $\hat{a}^{\dag}$, and a single pair of canonical operators $\hat{q}$ and $\hat{p}$.

The covariance matrix of a single-mode bosonic system prepared in a state $\rho$ is defined as~\cite{Ferraro2005}
\begin{equation}
\label{eq:def_covmat}
\sigma_{jk}[\rho] = \frac{1}{2} \langle \{ \hat R_j,\hat R_k \} \rangle_\rho - \langle \hat R_j \rangle_\rho \langle \hat R_k \rangle_\rho,
\end{equation}
where $\hat{\vec{R}}=(\hat{x},\hat{p})^T$  is the vector of single-mode quadrature operators $\hat x$ and $\hat p$, and the subscript implies that expectation values are calculated over state $\rho$. We also define the displacement vector $\vec{\bar{X}}[\rho]$ with components $X_k[\rho] = \langle \hat R_k \rangle_\rho$. %We also define the mean vector $\vec{\bar{X}}$, its components are $X_k = \langle R_k \rangle$.
A Gaussian state has a Gaussian Wigner function. %and therefore can be written in the following manner
%\begin{equation}
%W(\vec{X}) = \frac{1}{2\pi \sqrt{\det[\sigma]}} \exp \left[ -\frac{1}{2} \left(\vec{X} - \vec{\bar{X}} \right)^T \sigma^{-1}  \left(\vec{X} - \vec{\bar{X}} \right) \right],
%\end{equation}
%where $\vec{X}=(\operatorname{Re}z,\operatorname{Im}z)$. This definition
%means that, even if Gaussian states are continuous variable states, they
%are the simplest ones because they are fully determined by the knowledge
%of $\vec{\bar{X}}$ and $\sigma$.

To quantify non-Gaussianity of a generic state $\rho$, a reference Gaussian state $\tau$ should be defined. This is identified as the Gaussian state having the same covariance matrix and displacement vector  as $\rho$. That is
\begin{equation}
\label{eq:RefGaussianState}
\vec{\bar X}[\tau]= \vec{\bar X}[\rho], \qquad\sigma[\tau]=\sigma[\rho].
\end{equation}

Non-Gaussianity can now be defined as the distance 
between $\rho$ and $\tau$ calculated using, for instance, %an appropriate 
%metric, or by using 
the quantum relative entropy
\begin{equation}
S\left( \rho \middle\| \tau \right) = \Tr \left[ \rho ( \ln \rho - \ln \tau ) \right].
\end{equation}
%it can be proved that $0 \leq S\left( \rho_1 \middle\| \rho_2 \right) <
%\infty $, when it is properly defined, which means that the support of
%$\rho_1$ is contained in the support of $\rho_2$. 
We have that $S\left( \rho \middle\| \tau \right)=0$ iff
$\rho = \tau$. Although $S\left( \rho \middle\| \tau \right)$ is not symmetric in its arguments, and thus does not embody a proper metric,
it has been used widely to quantify the distinguishability of two states. %Indeed, the
%probability of mistaking $\rho_1$ for $\rho_2$ after $N\gg1$ measurements is proportional to $e^{-N S\left( \rho_1
%\middle\| \rho_2 \right)}$. 
This leads to the definition of the entropic measure of non-Gaussianity
\begin{equation}
\delta_{\text{E}} (\rho) =  S\left( \rho \middle\| \tau \right) =  \Tr \left[ \rho \ln \rho \right] - \Tr \left[ \rho \ln \tau \right] = S(\tau) - S(\rho),
\end{equation}
where $S$ denotes the von Neumann entropy and, owing the way $\tau$ is defined, we have that $S(\tau)=-\Tr \left[ \tau \ln \tau \right] = -\Tr \left[ \rho \ln \tau \right]$.
This measure satisfies a series of quite useful properties~\cite{Genoni2010}: it is additive under the tensor product
operation, and % Another important property is that $\delta_{\text{E}}(\rho)$ 
invariant under symplectic transformations, which are both very useful for the sake of our analysis. %, which means
%transformations generated by Hamiltonians quadratic or linear in the
%creation and destruction operators (which are known to preserve the
%Gaussian character of a state).

The von Neumann entropy of a single mode Gaussian state takes the very simple form
\begin{equation}
S(\rho_{\text{G}}) = h\left( \sqrt{\det \vec{\sigma }} \right),
\end{equation} 
where $h(x)=(x+\frac{1}{2}) \ln (x+\frac{1}{2}) - (x-\frac{1}{2}) \ln (x-\frac{1}{2})$. Thanks to this form, the entropic non-Gaussianity becomes
\begin{equation}
\delta_{\text{E}} (\rho) =  h\left( \sqrt{\det \vec{\sigma }} \right) - S(\rho),
\end{equation}
which is further simplified for pure states, as $S(\rho)=0$.

\subsubsection{Nonlinearity of a Potential}
%The ground state wave function of a quantum harmonic oscillator is a Gaussian 
%$\psi_{\text{H}} (x) = \left( \frac{\omega}{\pi}  \right)^{\frac{1}{4}} e^{-\frac{1}{2} \omega x^2}$, 
%where the mass and the Planck constant have been rescaled to unity. 
%This state is completely specified by the value of its frequency $\omega$ and also has a Gaussian Wigner function.

We consider a generic potential $V(x)$ and denote with $| \phi \rangle$ the ground state of the corresponding Hamiltonian. The first idea to quantify nonlinearity would be again using the
geometrical distance between the ground state of the potential and a
reference harmonic state, in particular for this purpose the Bures
metric has also been employed \cite{Paris2014}. This way of reasoning
has a downside because we have to choose a value for the frequency
$\omega$ of the reference harmonic oscillator. The most natural choice
is expanding the potential near its minimum and finding $\omega$ as a
function of the nonlinear parameters of the potential. However, 
determining this frequency is not always straightforward and for some
potentials exhbiting more than one minimum it may be even be misleading.

Instead of using a metric we choose to quantify nonlinearity using the entropic non-Gaussianity $\delta_{\text{E}}$, so that the measure of nonlinearity is defined as
\begin{equation}
\label{eq:DefGaussianNonLin}
\eta_{\text{NG}} [V]= \delta_{\text{E}} \left( |\phi\rangle \langle \phi | \right) = h \left(\sqrt{\det \vec{\sigma}}\right),
\end{equation}
this equality holds because the ground state is pure and $\vec{\sigma}$ is the covariance matrix of the ground state (we drop the dependence from the state when obvious).

This definition is more appealing than a geometric one because it does not require the determination of a reference potential for $V(x)$, but just the reference Gaussian state for the ground state of $V(x)$. This makes $\eta_{\text{NG}}$ independent of the specific features of the potential, since we do not need to know the behavior of $V(x)$ near its minimum to compute the reference frequency. 

Moreover, $\eta_{\text{NG}}$ inherits the property of the non-Gaussianity measure and is invariant under symplectic transformations~\cite{Ferraro2005}. This means that $\delta_{\text{NG}}$ assigns the same nonlinearity to oscillators which are displaced, rotated in phase space or squeezed, which is a reasonable property for a measure of nonlinearity.

\section{Exactly Solvable Nonlinear Oscillators}
\label{sec:potentials}
We now analyze quantitatively the relation between the figures of merit introduced in the previous Section, considering three exactly solvable anharmonic oscillators.
\subsection{Modified Harmonic Oscillator}
\begin{figure}[b]
	\centering
	\includegraphics[width=.5\textwidth]{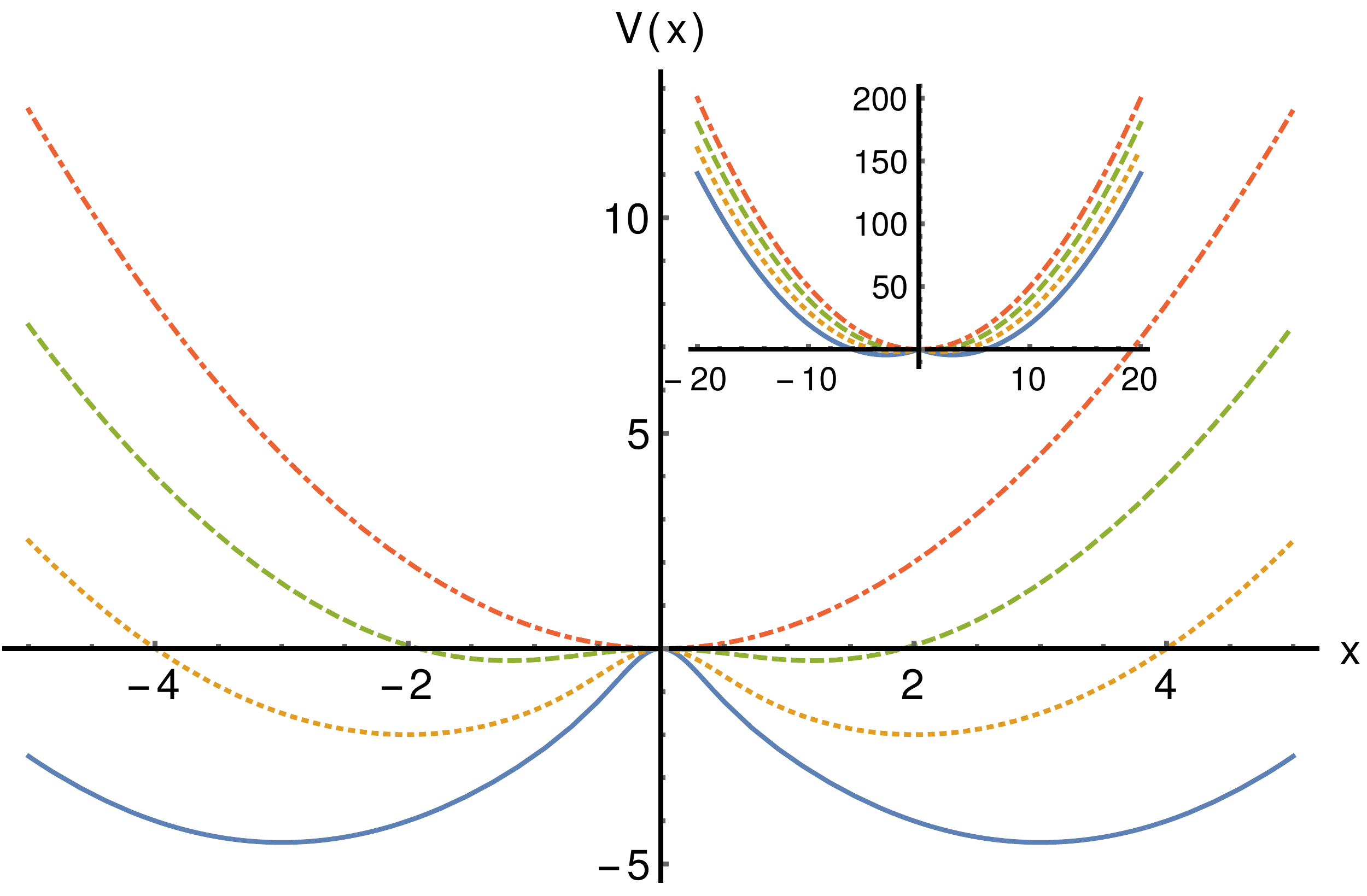}
	\caption{(Color online) The MHO potential $\alpha=1$ and $\beta = 3$ (solid blue), $2$ (dotted yellow), $1$ (dashed green) and the harmonic potential with unitary frequency and mass (dot-dashed orange). The inset represent the same graph at a larger scale, where we see the resemblance to the harmonic potential.}
	\label{fig:MHOpot}
\end{figure}

The Modified Harmonic Oscillator (MHO) potential is defined as~\cite{Bund2000} (throughout this manuscript we choose units such that $\hbar = m= 1$) 
\begin{equation}
\label{eq:MHOpot}
V_{\text{MHO}}(x) = \frac{\alpha ^2 x^2}{2}- \alpha \beta  x \tanh (\beta  x).
\end{equation}
Here $\alpha$ is a parameter corresponding to the frequency of the unmodified harmonic oscillator, while $\beta$ determines the deformation of the harmonic potential. The effects of this parameter on the shape of the potential is appreciated from Fig.~\ref{fig:MHOpot}, where $V_{\text{MHO}}(x)$ is plotted at a set value of $\alpha$ for different choices of $\beta$, showing that an increasing deformation parameter transforms a harmonic potential into a double-well one whose well-depth and separation both increase with $\beta$. 

The normalized wave-function of the ground-state of this potential can be found to read~\cite{Bund2000}
\begin{equation}
\label{eq:MHOground}
\phi_{\text{MHO}}(x) = \frac{\sqrt{2} e^{-\frac{1}{2} \alpha  x^2} \cosh (\beta  x)}{\sqrt[4]{\frac{\pi}{\alpha}} \sqrt{1+\exp[{\beta^2}/{\alpha }]}}.
\end{equation}
The associated energy is $E_0 = (\alpha - \beta^2)/{2}$. The covariance matrix of such least-energy state can be computed straightforwardly to be
\begin{equation}
\label{eq:MHOcorrmat}
 \vec{\sigma}^{\text{MHO}} = 
\begin{pmatrix}
 \frac{1}{2\alpha}+\frac{\beta^2}{\alpha^2} \frac{\exp[{\beta^2}/\alpha]}{1+\exp[{\beta^2}/\alpha]} & 0 \\
 0 & \frac{\alpha }{2}-\frac{\beta ^2}{1+\exp[{\beta ^2}/{\alpha }]} 
\end{pmatrix}.
\end{equation}
Its determinant is 
\begin{equation}
\det{\vec{\sigma}^{\text{MHO}}} =\frac{1}{4}  - \frac{\tau^2}{2} \frac{\left(  2 \tau^2  e^{\tau^2}-  e^{2\tau^2}+1\right)}{(e^{\tau^2}+1)^2}
\end{equation}
with $\tau=\sqrt{{\beta^2}/{\alpha}}$. Such dependence on $\tau$, rather than $\alpha$ and $\beta$ independently, is common to $\eta_{\text{NG}} = h(\det{\sigma})$ and the measure of nonlinearity based on the Bures distance (for the latter, we should choose the unmodified harmonic oscillator with frequency $\alpha$ as a reference). Both measures of nonlinearity increase monotonically with $\tau$.

The Wigner function associated with $\phi_{\rm MHO}$ can be written in terms of the suitably rescaled phase-space variables $q = \beta x$ and $p = \frac{\beta}{\alpha} y$ as~\cite{Bund2000}
\begin{equation}
\label{eq:MHOwig}
W_{\text{MHO}}(q,p) = e^{-\frac{q^2+p^2}{\tau^2}} \frac{\cosh (2 q)+
e^{\tau^2}\cos (2p)}{\pi\tau^2(1+e^{\tau^2})},
%W_{\text{MHO}}(x,y) = \frac{e^{-\alpha  x^2-\frac{y^2}{\alpha 
%}} [e^{-\tau^2} \cosh (2 \beta  x)+
%\cos ({2 \beta  y}/{\alpha })]}{\pi  
%(1+e^{-\tau^2})}.
\end{equation}
which shows again the key role played by $\tau$ and, in turn, that the non-classicality measure based on the volume of the negative part of $W_{\rm MHO}(q,p)$ is  determined by such parameter. 
%By rescaling variables as In order to get the volume of the negative part of the function we have
%to integrate the absolute value of \eqref{eq:MHOwig} and by changing the
%variables to  the integral
%becomes \begin{equation}
%\label{eq:MHOintvariables}
%\begin{split}
%&\iint \! \mathrm{d} x \mathrm{d}y \, \left|W_{\text{MHO}}(x,y)\right| = \\
%&\frac{1}{\pi \tau^2  \left(1 + e^{\tau^2}\right)}\iint \! \mathrm{d}q 
%\mathrm{d}p \, \left\lvert  e^{-\frac{q^2 + p^2}{\tau^2}} 
%\left[ \cosh (2 q)+ e^{\tau^2} \cos \left(2 p\right)\right] \right\lvert,
%\end{split}
%\end{equation} 
%making evident that it only depends on the parameter $\tau$.

\begin{figure}[!t]
\centering
\includegraphics[width=.5\textwidth]{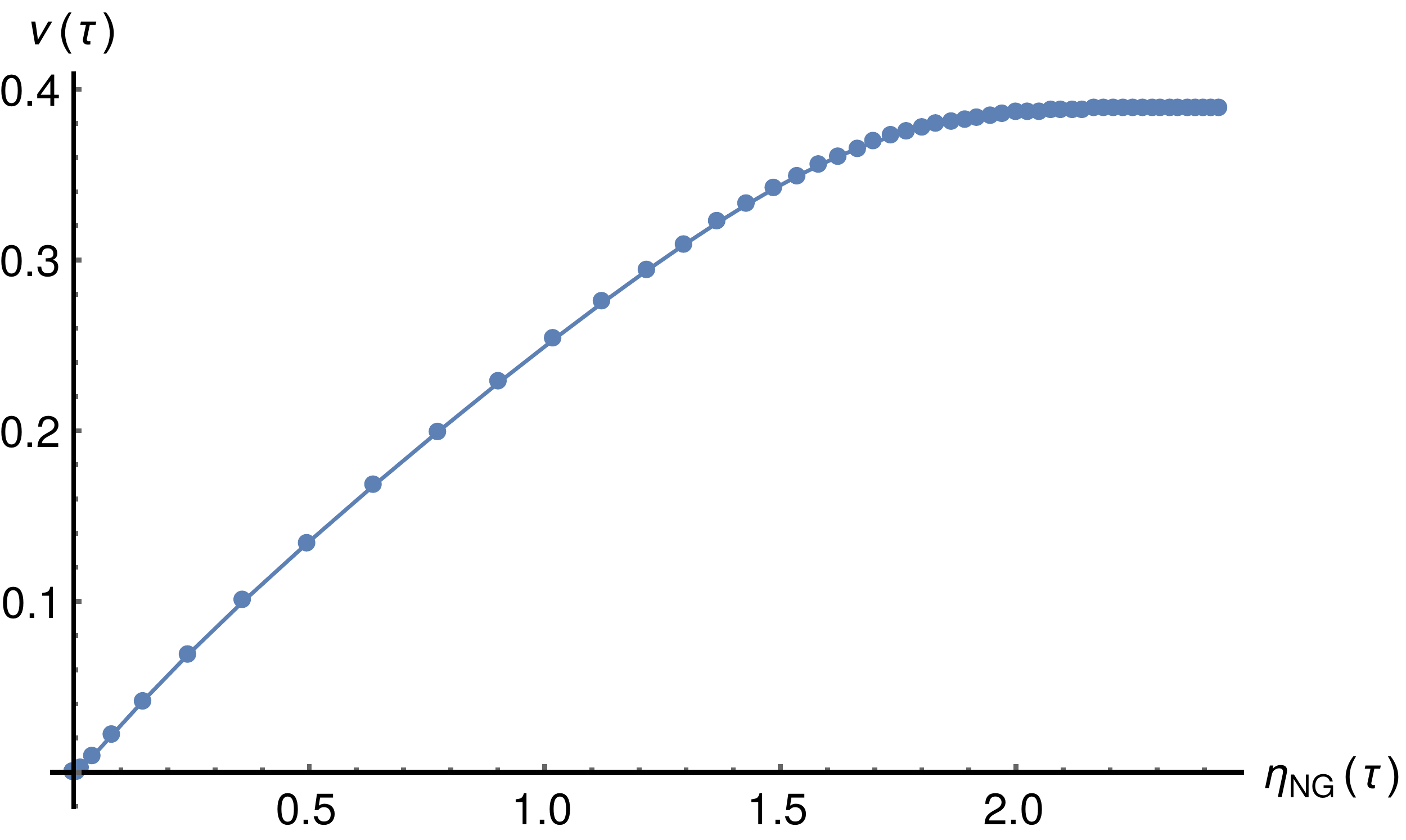}
\caption{(Color online) Parametric plot of the $W$-nonclassicality measure
$\nu(\tau)$ versus the degree of nonlinearity for the MHO potential and for $\tau\in[0.1,6]$.} \label{fig:MHOdelta}
\end{figure}

In order to understand how $W$-nonclassicality and nonlinearity are 
related to each other, we have studied both quantities against $\tau$. In Fig.~\ref{fig:MHOdelta} we report the resulting parametric plot, 
showing that $\nu$ monotonically increases with $\eta_{\text{NG}}$, thus
supporting the idea that a growing degree of anharmonicity of the potential results in increased nonclassicality of the corresponding ground state.

%\subsubsection{Entanglement potential}
However, the picture changes significantly as soon as we consider
P-nonclassicality quantified by the entanglement potential which, as
said, can single out more detailed features of quantumness. Indeed, at
variance with what has been found above, such figure of merit turns out
to depend on $\alpha$  and $\beta$ independently. The reason for such a
difference in behavior should be ascribed to the fact that entanglement
at the output of a beam splitter can be originated either by a
non-Gaussian input state or by Gaussian single-mode squeezing.
In other words, nonlinearity is needed to generate
W-nonclassicality, while P-nonclassicality may be obtained using just
squeezing.

\begin{figure}[t!]
\centering
{\bf (a)}\hskip0.5\columnwidth{\bf (b)}\\
\includegraphics[width=.49\columnwidth]{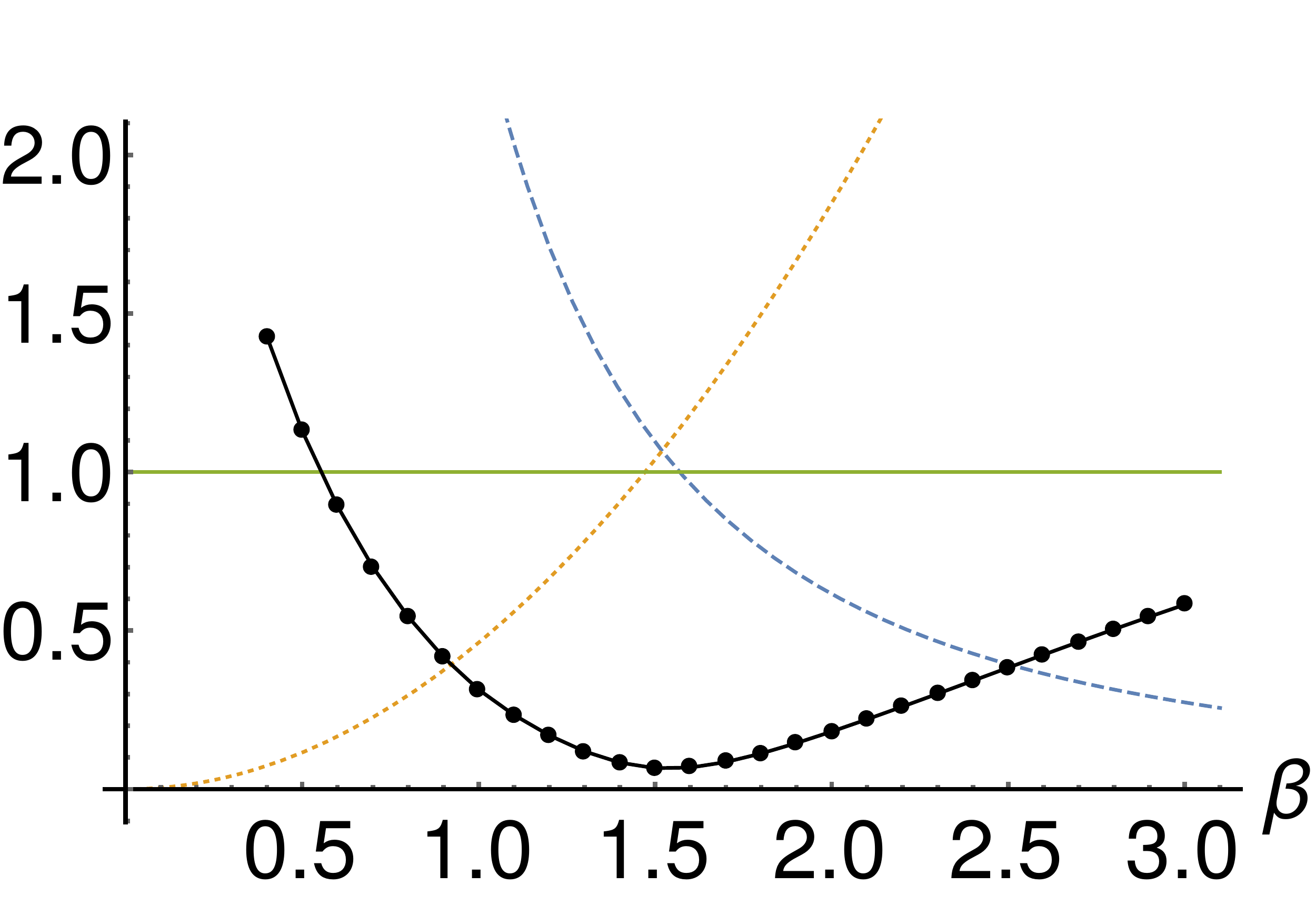}
\includegraphics[width=.49\columnwidth]{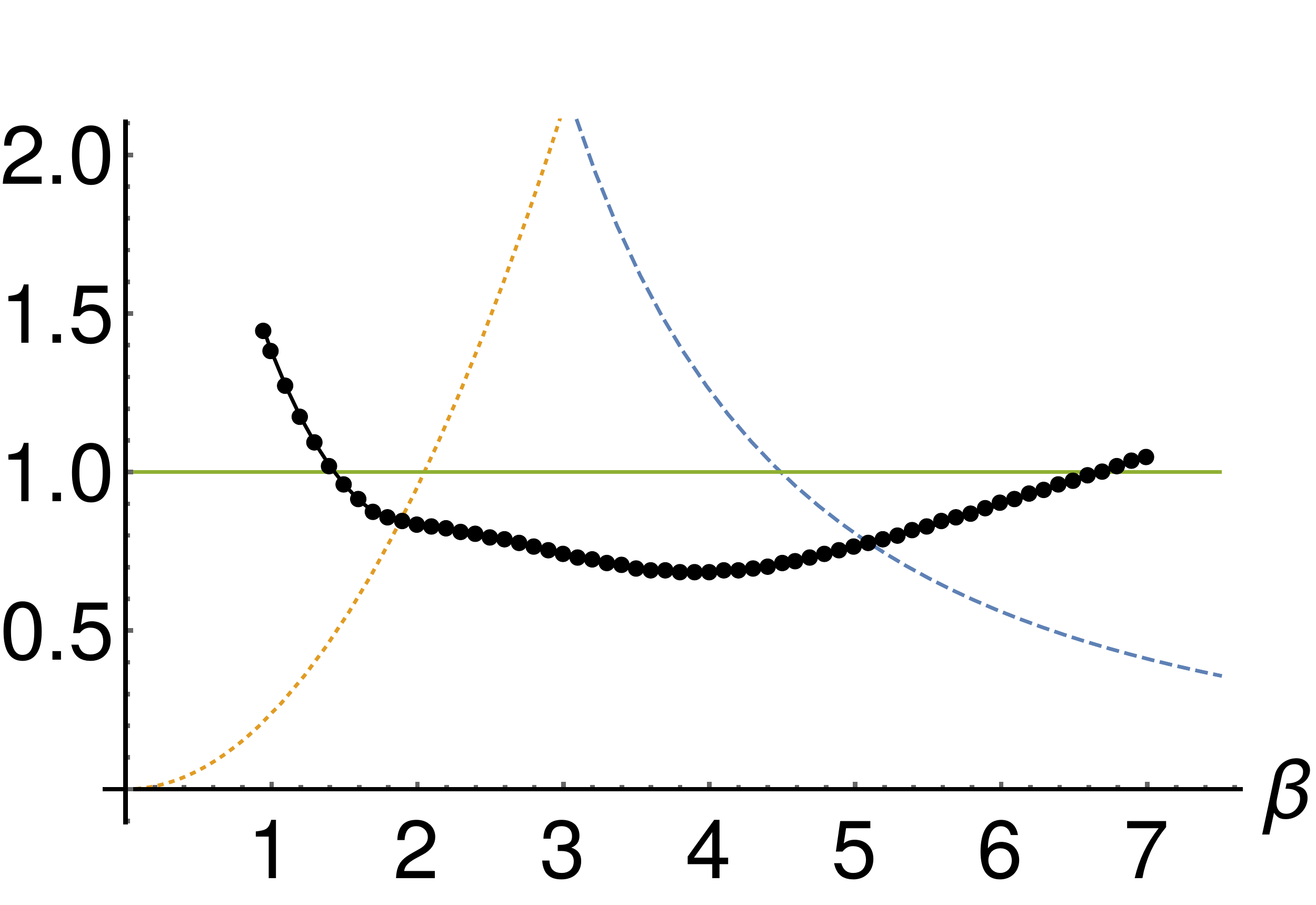}\\
{\bf (c)}\hskip0.5\columnwidth{\bf (d)}\\
\includegraphics[width=.49\columnwidth]{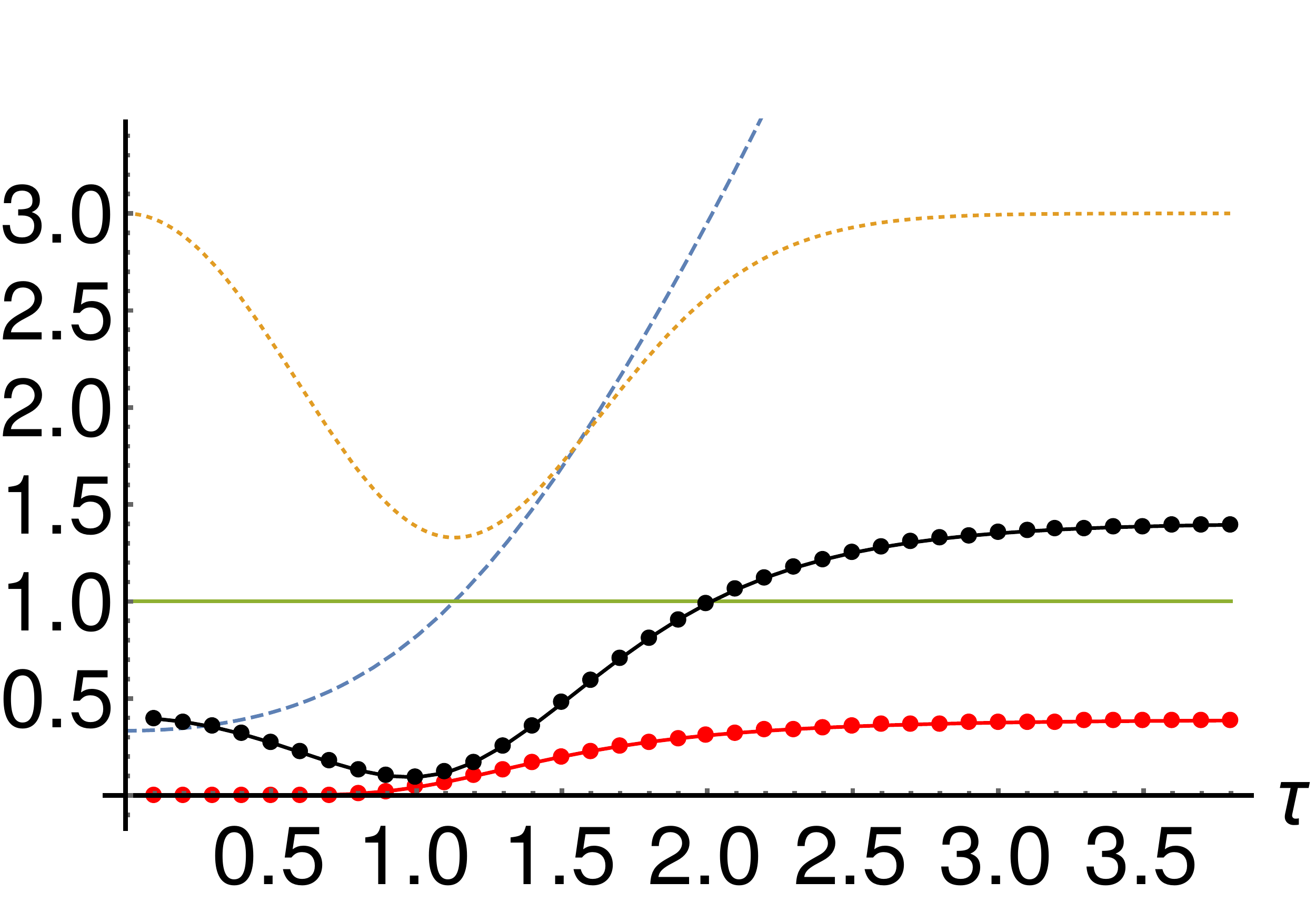}
\includegraphics[width=.49\columnwidth]{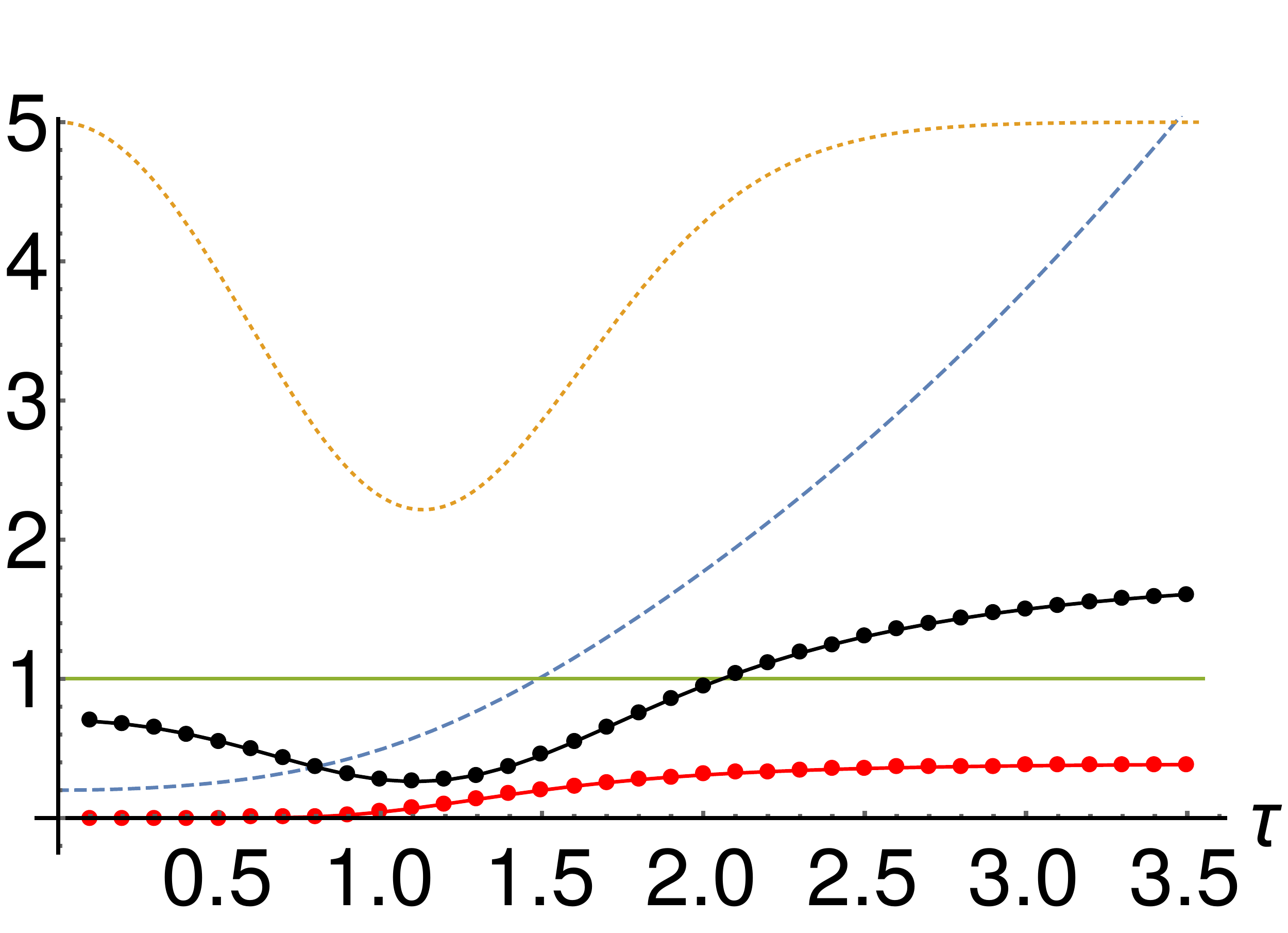}
\caption{(Color online) Entanglement potential and squeezing for the
MHO. In panels {\bf (a)} and {\bf (b)} [{\bf (c)} and {\bf (d)}] we plot $r_x$ (dashed blue curve), 
$r_p$ (dotted orange curve), the entanglement potential (P-nonclassicality)
$\ep$ (black dots), and $W$-nonclassicality $\nu$ (red dots) against $\beta$ [$\tau$] for $\tau=1$ and $\tau=3$ [$\alpha=3$ and $\alpha=5$]. Squeezing is observed for either $r_x<1$ or $r_p<1$ (i.e. variances of the perturbed ground state below the values of the vacuum state of a harmonic oscillator).}
\label{fig:MHOsqueezingvsEP}
\end{figure}

In order to illustrate this clearly, in Fig.~\ref{fig:MHOsqueezingvsEP} we show the entanglement
potential and squeezing for the MHO both as a function of $\beta$ for fixed
values of $\tau$, and as a function of $\tau$ at set values of
$\alpha$. %At set values of $\tau$, $\nu$ is constant we don't show the $W$-nonclassicality
%$\nu$ because it is constant (as well as the nonlinearity).
The squeezing  in Fig.~\ref{fig:MHOsqueezingvsEP} is shown in terms of 
the ratios 
\begin{equation}
r_x=\frac{\sigma^{\rm MHO}_{11}}{\sigma^0_{11}}=2\sigma^{\rm MHO}_{11},\quad r_p=\frac{\sigma^{\rm MHO}_{22}}{\sigma^0_{22}}=2\sigma^{\rm MHO}_{22}
\end{equation}
 with $\sigma^{0}_{11}=\sigma^{0}_{22}=1/2$ the variances of position and momentum calculated over the vacuum state of the harmonic potential. 
%$\overline{\Delta p}$ on the ground state to the fluctuations of 
%the harmonic vacuum $\overline{\Delta x_0} = 
%\overline{\Delta p_0} = 1/\sqrt{2}$, i.e. 
%\begin{equation}
%r_x^2 = \frac{\overline{\Delta x}^2}{\overline{\Delta x_0}^2} = 2
%\overline{\Delta x}^2 \qquad 	r_p^2 = \frac{\overline{\Delta
%p}^2}{\overline{\Delta p_0}^2} = 2 \overline{\Delta p}^2\,. 
%\end{equation}
Squeezing is found in the ground state of the MHO for either $r_x<1$ or $r_p<1$. 
As it is apparent from Fig.~\ref{fig:MHOsqueezingvsEP}, the behavior of $\ep$ is rather 
different from $\nu$, and its features may be understood looking at 
squeezing. In particular, we see that $\ep$ grows when 
the ground state exhibits squeezing.

\subsection{Morse potential}
\begin{figure}[!t]
	\centering
	\includegraphics[width=.5\textwidth]{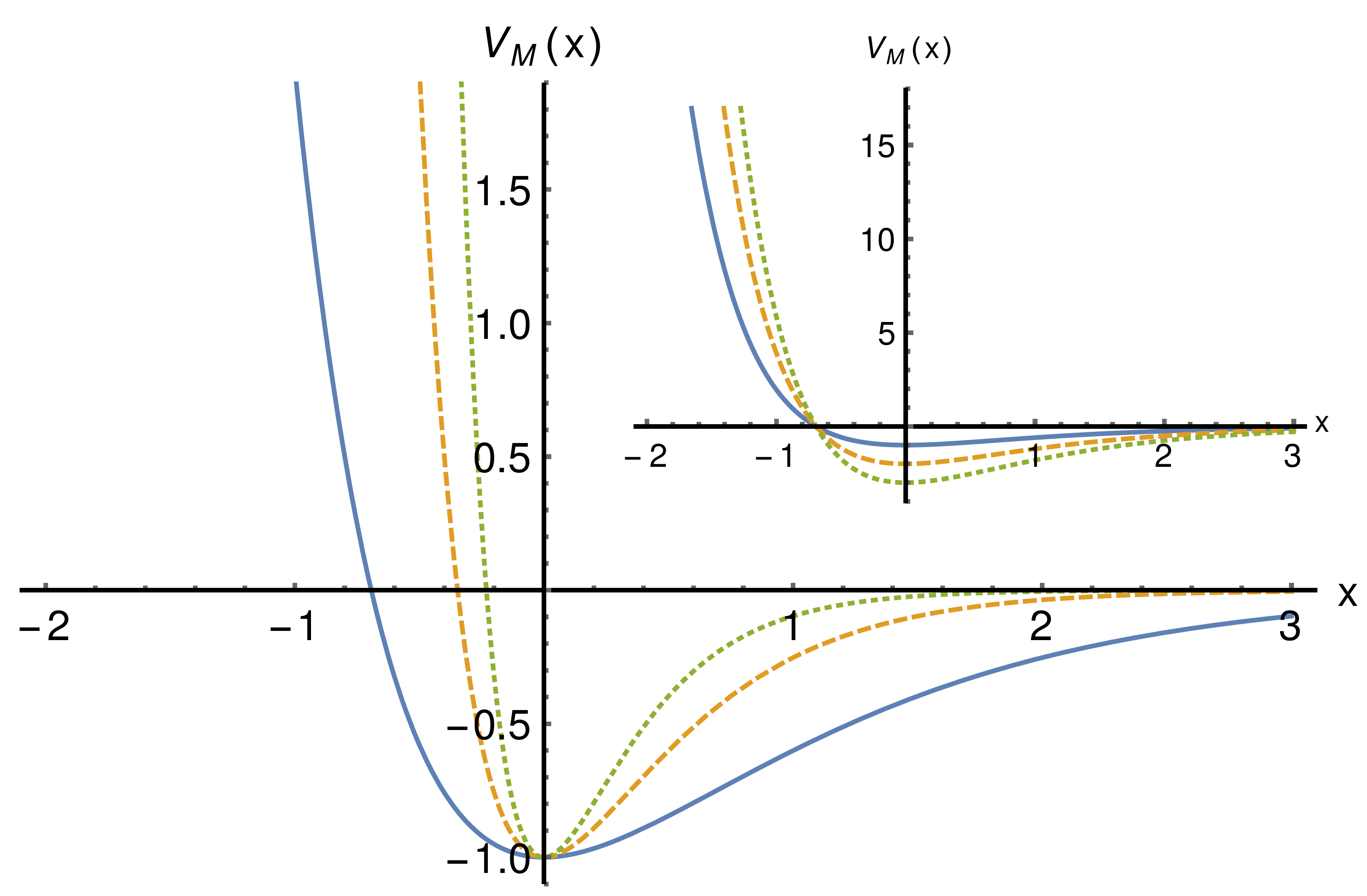}
	\caption{(Color online) The Morse potential $V_{\rm M}(x)$ for $D = 1$ and $\alpha = 1$ (solid blue), $2$ (dashed orange), $3$ (dotted green). The inset shows the potential for $\alpha=1$ and $D = 1$ (solid blue), $2$ (dashed orange), $3$ (dotted green).}
	\label{fig:MorsePot}
\end{figure}

The Morse potential has been introduced as an
approximation to the potential energy of diatomic molecules as it
provides a better description of the vibrational structure than the
(quantum) harmonic oscillator~\cite{Morse1929}. The form of the potential is 
\begin{equation}
\label{eq:MorsePot}
V_{\text{M}} = D \left( e^{-2 \alpha x} - 2 e^{-\alpha x} \right),
\end{equation}
where $x$ is the distance from the minimum of the
potential, the parameter $D > 0$ determines the depth of the well, while
$\alpha$ controls its width. Expanding the two exponentials for $\alpha
\to 0$ at fixed $D$ we get the harmonic limit, which is an oscillator
with frequency $\omega_{\text{M}} = \sqrt{2D} \alpha$. The
potential is plotted in Fig.~\ref{fig:MorsePot} for different values of the
parameters.

The  Schr\"odinger equation associated with this potential
can be solved analytically, the energy eigenvectors being labelled by two
quantum numbers, which we label here $N$ and $\nu$. The first is related to the parameters
of the potential as $N = -{1}/{2} +
{\sqrt{2D}}/{\alpha}$. The second, which can take values $\nu=0,1,2,..,\lfloor{N}\rfloor$, counts the number of anharmonic excitations of the system. As we want at least one bound state, we require $N>0$. We thus have the constraint $\alpha < 2 \sqrt{2 D}$. The limiting case where we
have just one bound state (the ground state) is achieved for %$D \to 0$ or 
$\alpha \to 2\sqrt{2D}$. The wave-function of the ground state is
\begin{equation}
\label{eq:MorseGS}
\phi_{\text{M}} (x) = (2 N + 1)^N \sqrt{\frac{\alpha}{(N-1)!}} e^{-\alpha x N - (N + \frac{1}{2})e^{-\alpha x}}
\end{equation} 
with associated energy $E=-\alpha N^2/2$. % ($N$ can be real valued and $N!$ means $\Gamma(N+1)$).
The behavior of the nonlinearity of the Morse potential can be understood by looking at the form of
the potential in Fig.~\ref{fig:MorsePot}, as opposed to the harmonic
one~\cite{Paris2014}: For any fixed value of $D$ [$\alpha$] we expect an increase [decrease] of nonlinearity 
for increasing $\alpha$ [$D$]. 

The covariance matrix associated with the ground state in Eq.~\eqref{eq:MorseGS} is
\begin{equation}
\label{eq:MorseCorMat} \vec{\sigma}^{\text{M}} =
\begin{pmatrix}
\frac{\psi ^{(1)}(2 N)}{\alpha ^2} & 0 \\
0 &  \frac{\alpha ^2 N}{2} 
\end{pmatrix},
\end{equation}
where $\psi^{(n)}(z)$ is the polygamma function $\psi^{(n)}(z) =
\frac{\mathrm{d}^{n+1}}{\mathrm{d}z^{n+1}} \log \Gamma (z)$, $\Gamma(Z)$
being the Euler Gamma function.
The determinant of this correlation matrix, and thus the Bures distance
from the reference harmonic oscillator, depend just on $N$ or, equivalently, on the combination
${\sqrt{2D}}/{\alpha}$. In this case both measures of nonlinearity are monotonically
decreasing functions of $N$.

\begin{figure}[!h]
\centering
\includegraphics[width=.5\textwidth]{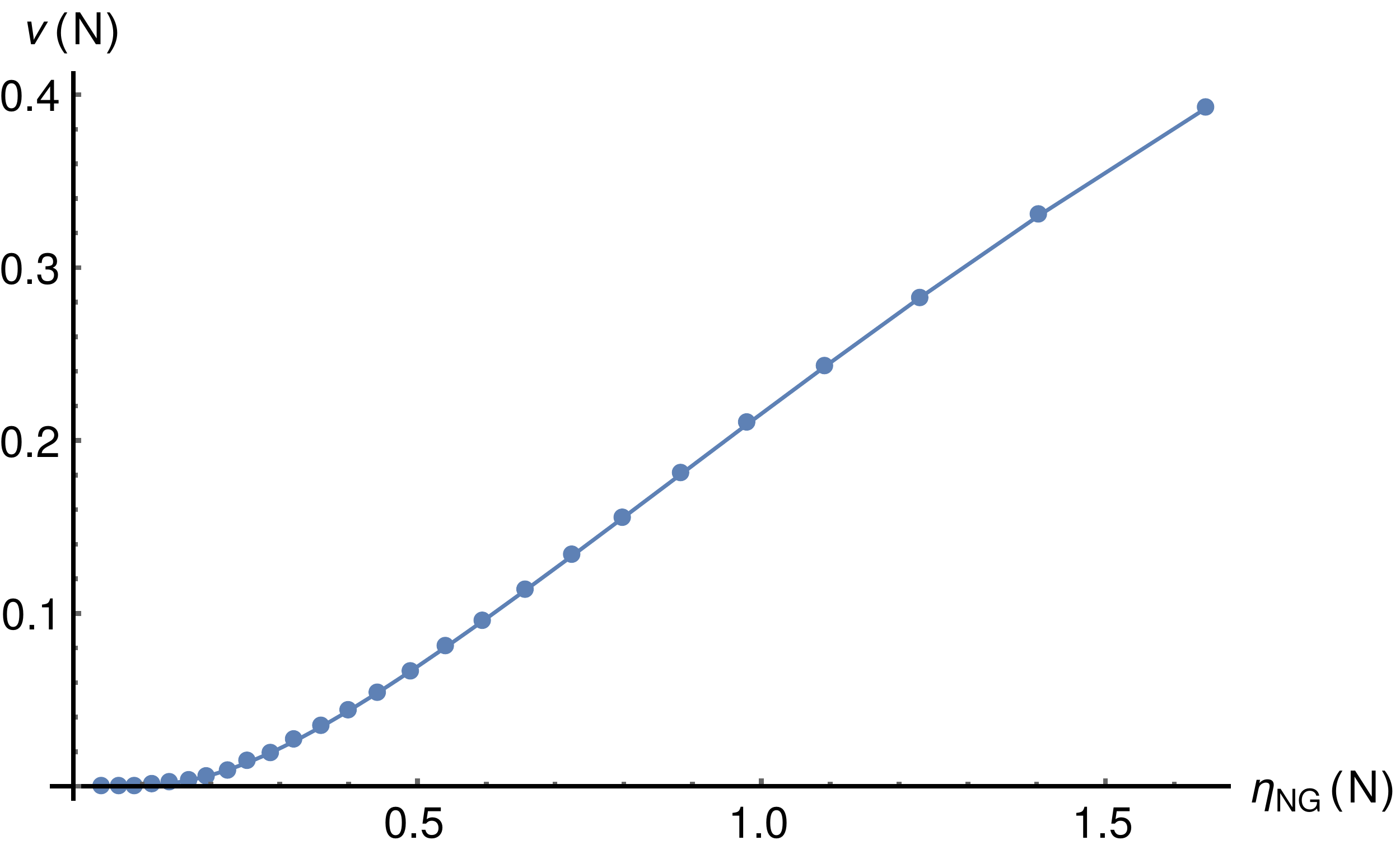}
\caption{(Color online) Parametric plot of the $W$-nonclassicality $\nu$
versus the degree of nonlinearity $\eta$ for a Morse potential with $D=1$ and
$\alpha\in[0.15,2.7]$, i.e. $N\in[0.0238,8.928]$.}
\label{fig:Morsenu}
\end{figure}

The Wigner function for the ground state of the Morse potential reads as follows~\cite{Frank2000}
\begin{equation}
W_{\text{M}}(x,p)= \frac{2(2N + 1 )^{2N}}{\pi \Gamma(2N)} e^{-2N\alpha x} K_{-2\I p / \alpha}\left( \left(2N + 1\right) e^{-\alpha x} \right),
\end{equation}
where $K_{\gamma}(z)$ is the Macdonald function of (non-integer) order $\gamma$. In order to calculate the measure of nonclassicality $\nu$, we rescale the phase-space variables to $q=\alpha x$ and $p=\frac{y}{\alpha}$, and evaluate 
\begin{equation}
\label{eq:MorseWigInt}	
\begin{split}
&\iint \! \mathrm{d}x\, \mathrm{d}y \, \left\lvert W_{\text{M}}(x,y) \right\lvert =\\
&\iint \! \mathrm{d}q\, \mathrm{d}p \left\lvert \frac{2e^{-2N q}}{\pi \Gamma(2N)} (2N + 1 )^{2N}  K_{-2\I p}\left( \left(2N + 1\right) e^{-q} \right)  \right\lvert,
\end{split}
\end{equation}
which shows that the only relevant parameter is $N$. The numerical integration of Eq.~\eqref{eq:MorseWigInt} is
challenging and was carried out with the aid of the CUBA
libraries~\cite{Hahn2005}.
The degree of $W$-nonclassicality $\nu$ is found to monotonically decreases 
with $N$, and the
parametric plot of nonclassicality versus nonlinearity in
Fig.~\ref{fig:Morsenu} reveals a monotonic behavior, strengthening the link between such features and reinforcing the idea that nonlinearity might play the role of a catalyst for nonclassicality. 

\begin{figure}[!b]
\centering
{\bf (a)}\hskip0.5\columnwidth{\bf (b)}\\
\includegraphics[width=.49\columnwidth]{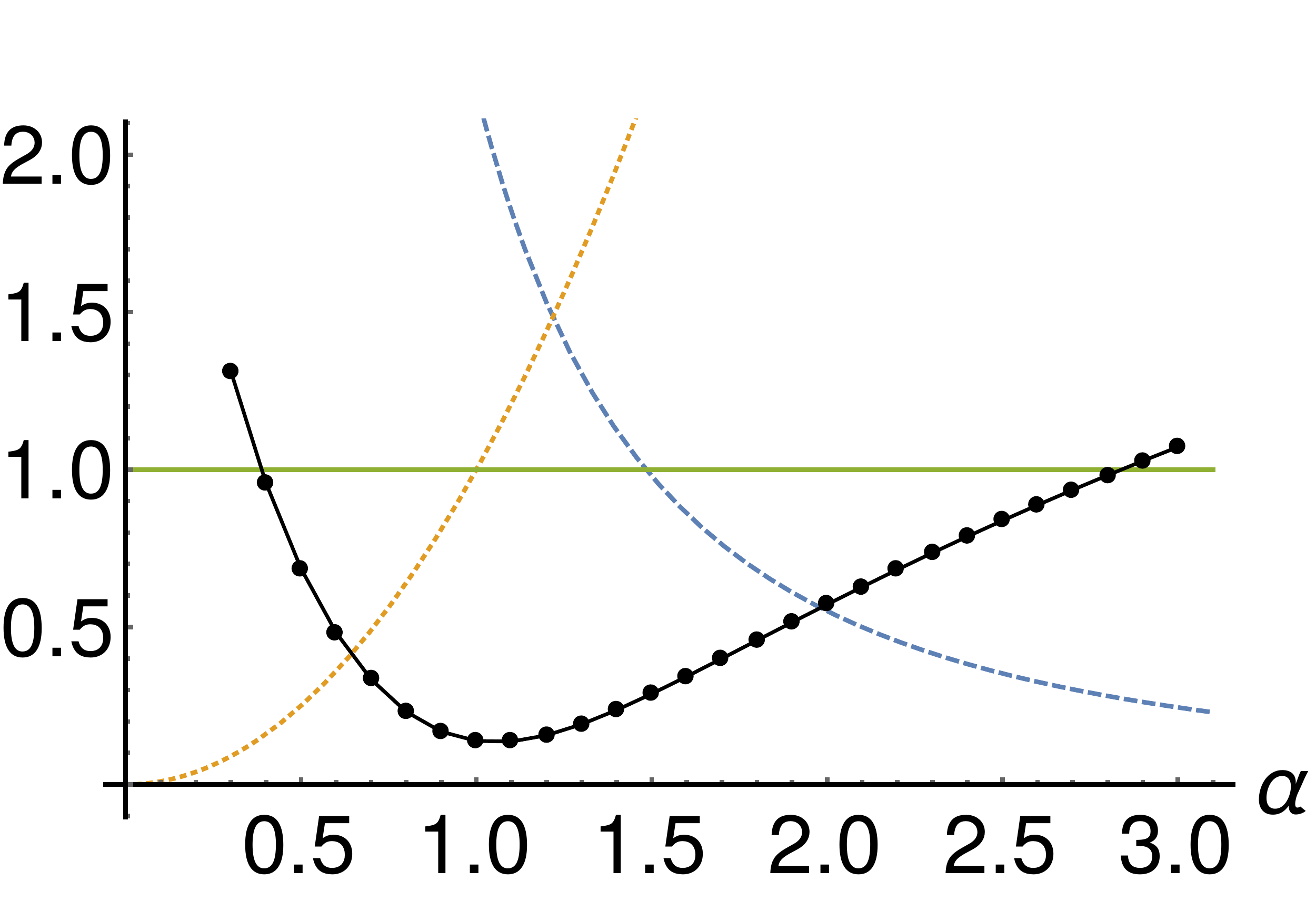}
\includegraphics[width=.49\columnwidth]{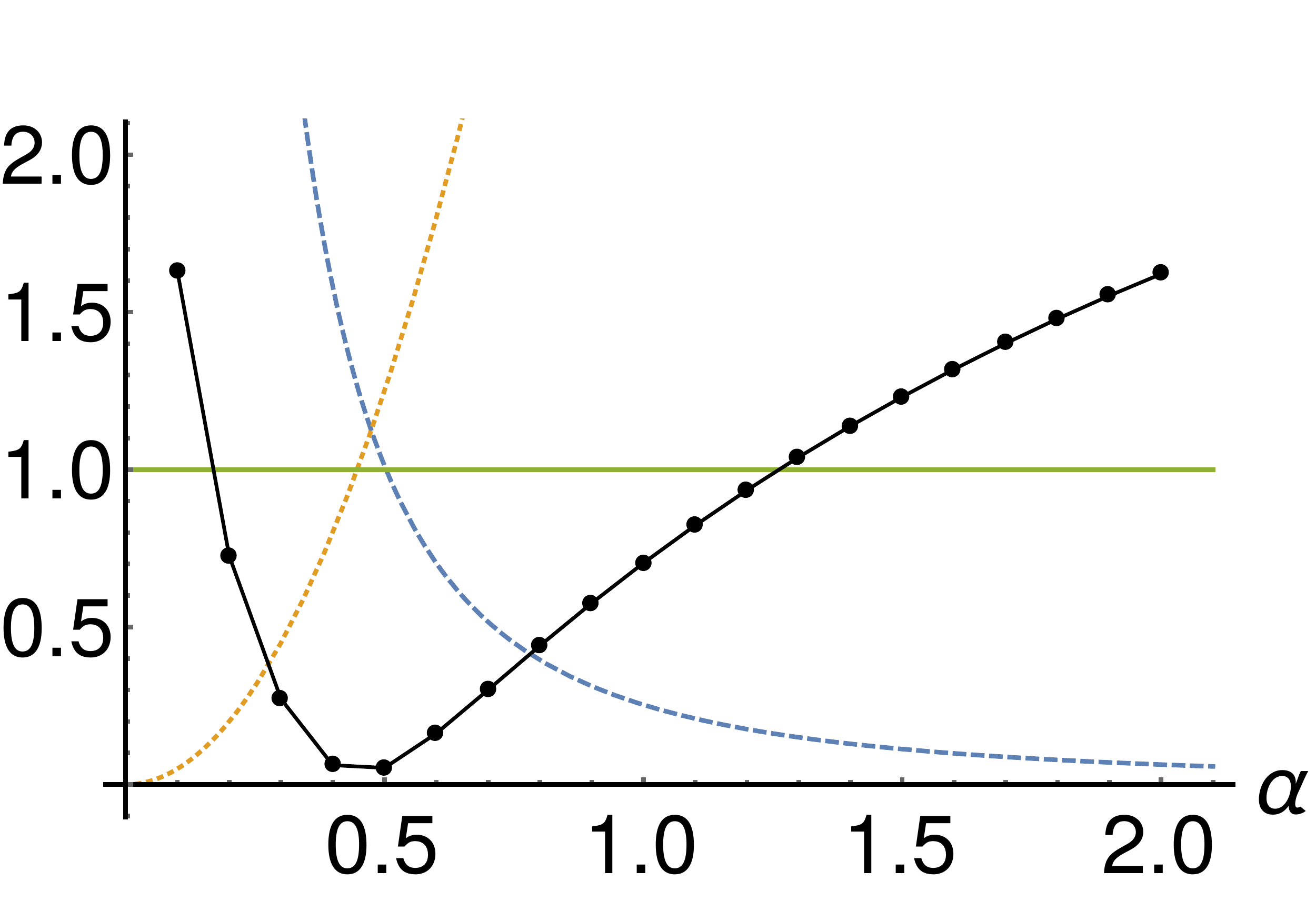}
{\bf (c)}\hskip0.5\columnwidth{\bf (d)}\\
\includegraphics[width=.49\columnwidth]{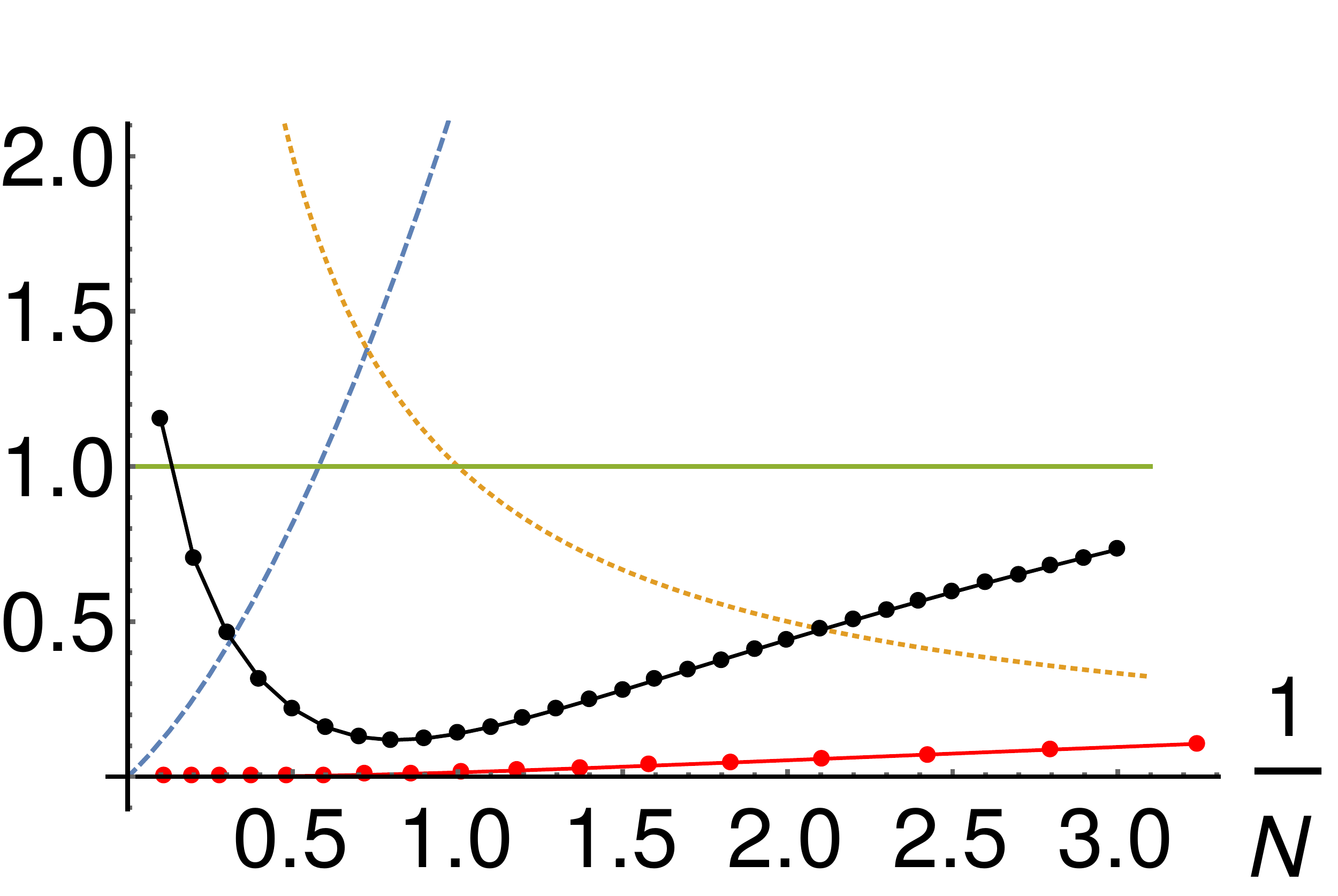}
\includegraphics[width=.49\columnwidth]{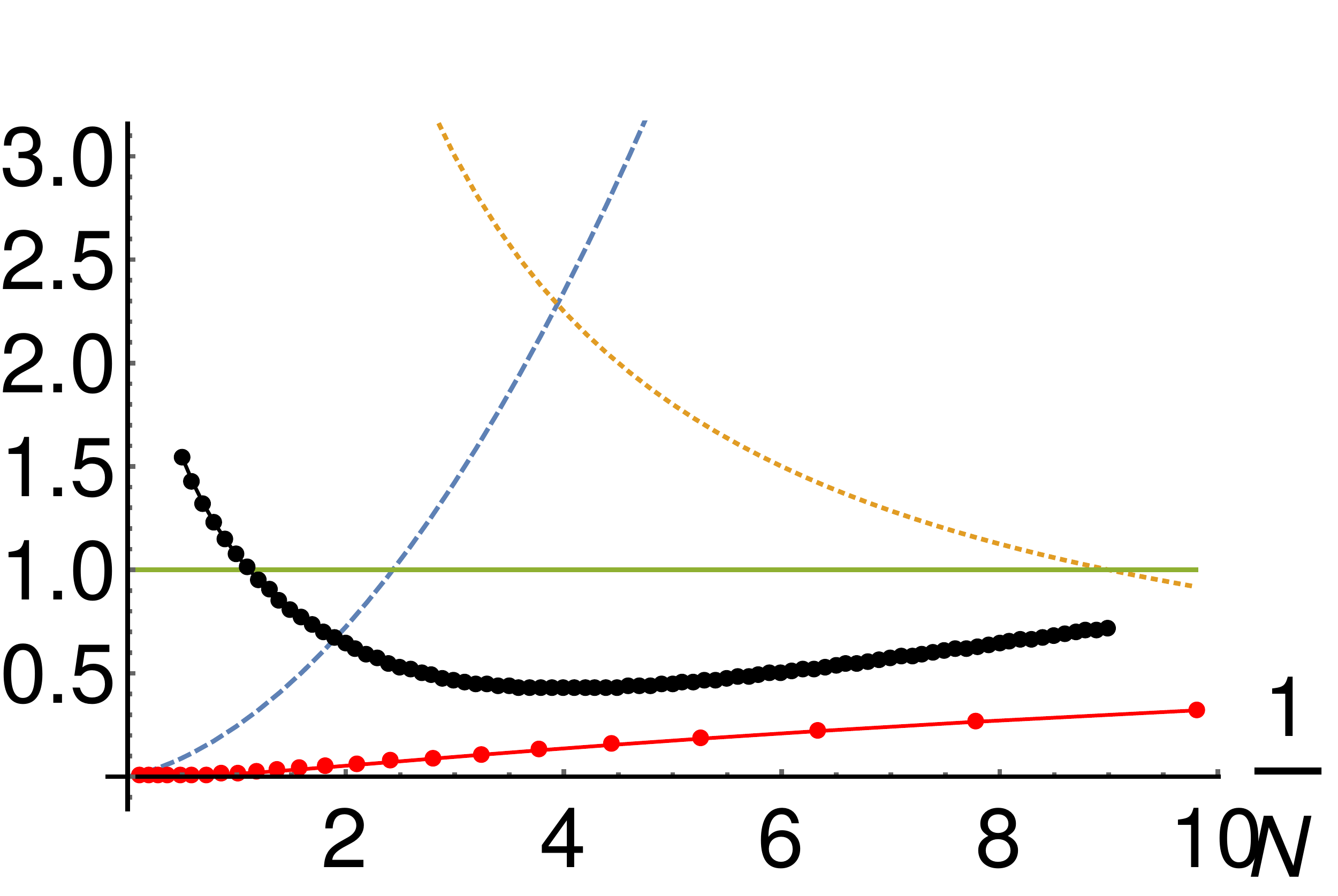}
\caption{(Color online) Entanglement potential and squeezing for
the Morse oscillator. In panels {\bf (a)} and {\bf (b)} [{\bf (c)} and {\bf (d)}] we plot $r_x$ (dashed blue curve), 
$r_p$ (dotted orange curve), the entanglement potential (P-nonclassicality)
$\ep$ (black dots), and $W$-nonclassicality $\nu$ (red dots) against $\alpha$ [$1/N$] for $N=1$ and $N=5$ [$\alpha=1$ and $\alpha=3$]. Squeezing is observed for either $r_x<1$ or $r_p<1$ (i.e. variances of the perturbed ground state below the values of the vacuum state of a harmonic oscillator).}
\label{fig:MorsesqueezingvsEP}
\end{figure}

The situation regarding the entanglement potential is completely analogous to what we found for the MHO, as it depends on both parameters.
In Fig.~\ref{fig:MorsesqueezingvsEP} we report the same kind of graphs, both with $N$ fixed and $\alpha$ fixed,
which show that the behavior of $\ep$ is explained by the squeezing of the state.

\subsection{Pöschl-Teller potential}
\begin{figure}[t!]
\centering
\includegraphics[width=.5\textwidth]{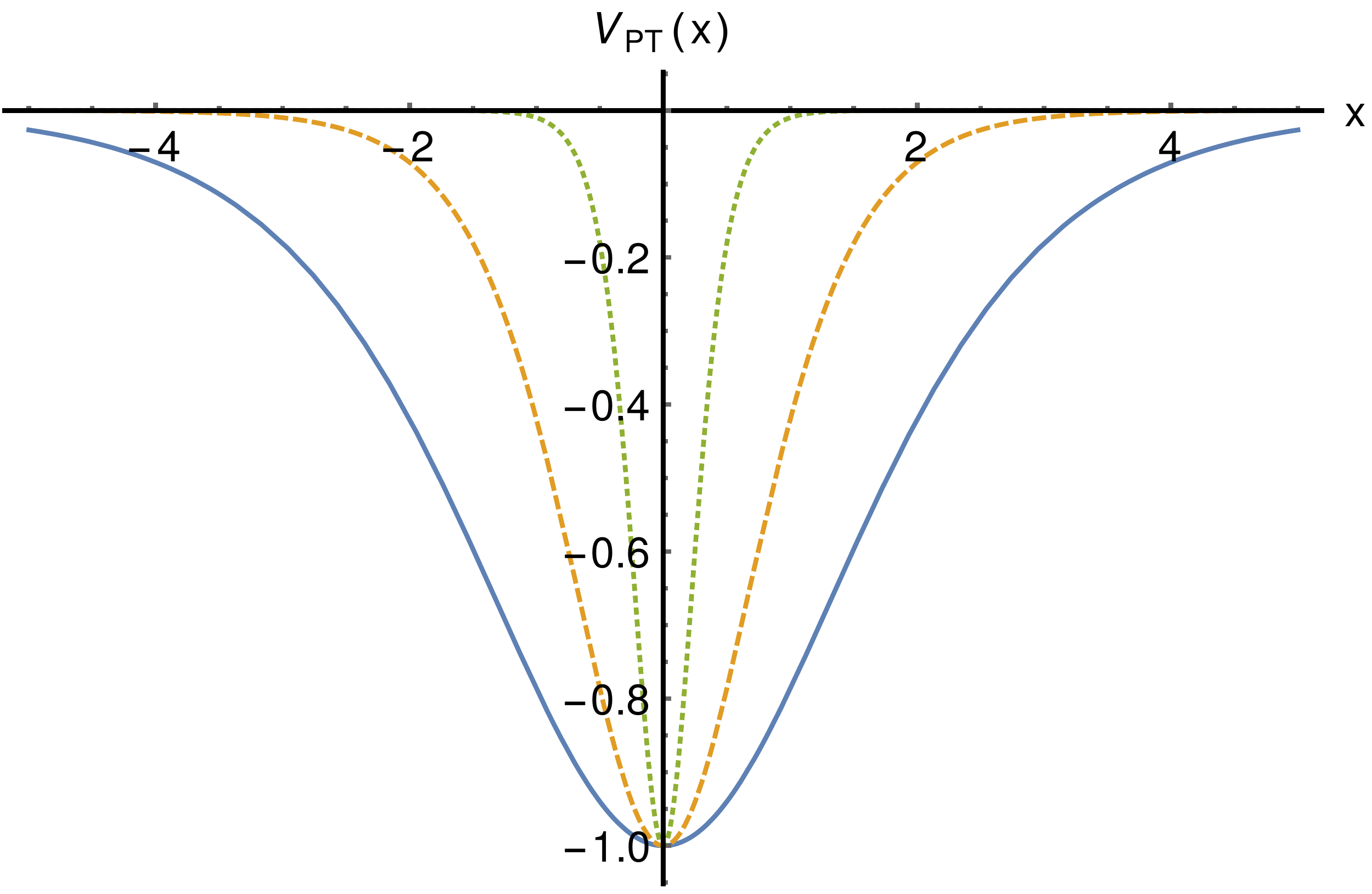}
\caption{(Color online) The Posh-Teller potential with $A_{\rm PT}=1$ and $\alpha=1/2$ (solid blue), $1$ (dotted orange) and $3$ (dashed green).}
	\label{fig:PTpot}
\end{figure}

The modified P\"oschl-Teller potential (PT) is defined as
\begin{equation}
\label{eq:PTpot}
V_{\text{PT}}(x)=-{A_{\rm PT}}{\cosh ^{-2}(\alpha  x)},
\end{equation}
where $A_{\rm PT}>0$ is the depth of the potential and $\alpha$ is
connected to its range. The harmonic limit is obtained at fixed $A_{\rm PT}$ for
$\alpha \to 0$ and the frequency of the reference harmonic oscillator is $\omega_{\rm PT} = \sqrt{2A_{\rm PT}}\alpha$. 
As for the Morse potential, we have a quantum number $s$ that labels the energy eigenstates and counts the anharmonic 
excitations. It is related to the parameters of the potential through the relation $A_{\rm PT}=\frac{1}{2} \alpha^2 s ( s + 1)$. Therefore, the 
request for the existence of at least one bound state translates into $s = \frac{1}{2}\left( -1 + \sqrt{1 + 8 A_{\rm PT} / \alpha^2} \right)
>0$. Fig.~\ref{fig:PTpot} shows the dependence of the PT potential on the position coordinate.

The ground state of  the system reads
\begin{equation}
\label{eq:PTground}
\phi_{\text{PT}}(x) = \frac{1}{\pi^{\frac{1}{4}} } \sqrt{\frac{\alpha  \Gamma \left(s+\frac{1}{2}\right)}{\Gamma (s)}} {\cosh^{-s}(\alpha  x)},
\end{equation}
with associated energy $E=-\alpha^2 s^2/2$.

Differently from the previous cases, the covariance matrix of the ground state is rather involved and will not
be reported here. In line with the case of the previous two anharmonic potentials studied here, its determinant depends only on $s$ (or, equivalently, on $A_{\rm PT}/{\alpha^2}$). Again, both $\eta_{\text{NG}}$ and the Bures nonlinearity are
monotonically decreasing function of $s$ only.

The Wigner function of state $\phi_{\rm PT}(x)$ in Eq.~\eqref{eq:PTground} is known
analytically for the case of $A_{\rm PT}=\alpha^2$ \cite{Bund2000}. In
this case, the measure $\nu$ is an $s$-dependent constant, as it can be seen by rescaling the relevant variables as $p'=\frac{p}{\alpha}$, $x'=\alpha x$, $y'=\alpha y$ and evaluating the integral %As
%a matter of fact we can prove that it also depends only on the parameter
%$s$ by appropriately changing the variables. If we apply the definition
%of Wigner function we have 
\begin{equation}
W_{\text{PT}}(x,p) = \int\! \mathrm{d}y  \, \phi^*_{\text{PT}}\left(x-\frac{y}{2}\right) \phi_{\text{PT}}\left(x+\frac{y}{2}\right) e^{-\I y p},
\end{equation}
which embodies the definition of Wigner function. 
%and to get $\nu$ we need the integration 
%\begin{equation}
%\iint \!  \mathrm{d}x \mathrm{d}p  \,\left\lvert W_{\text{PT}}(x,p) \right\lvert.
%\end{equation}
%If we change variables to $p'=\frac{p}{\alpha}$, $x'=\alpha x$ and
%$y'=\alpha y$ we can see that the dependency on $\alpha$ vanishes even
%without evaluating the integral. Unfortunately we do not have
%$W_{\text{PT}}$ in closed form, so we could not compute $\nu(s)$.

\begin{figure}[t]
\centering
{\bf (a)}\hskip0.5\columnwidth{\bf (b)}\\
\includegraphics[width=.49\columnwidth]{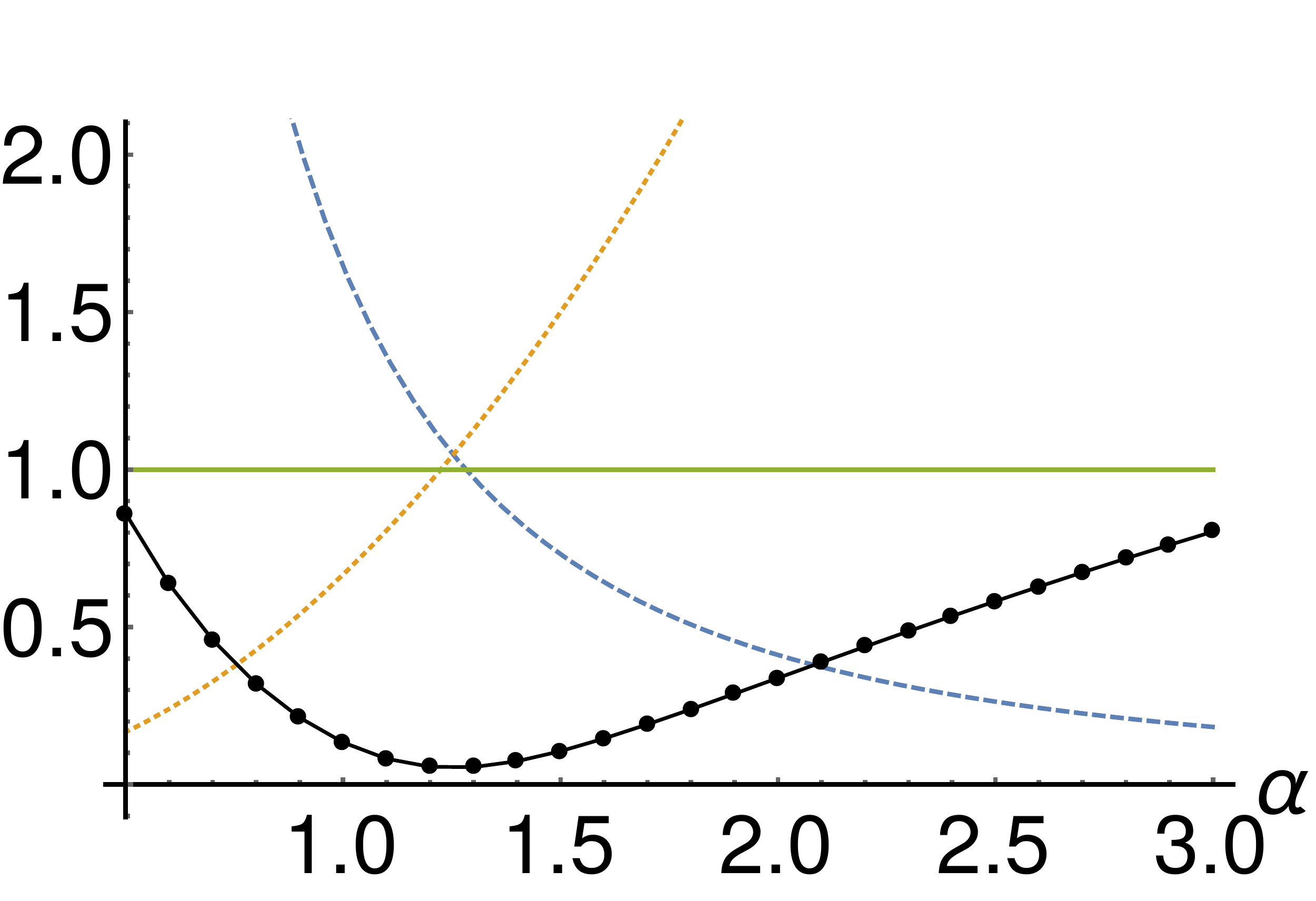}
\includegraphics[width=.49\columnwidth]{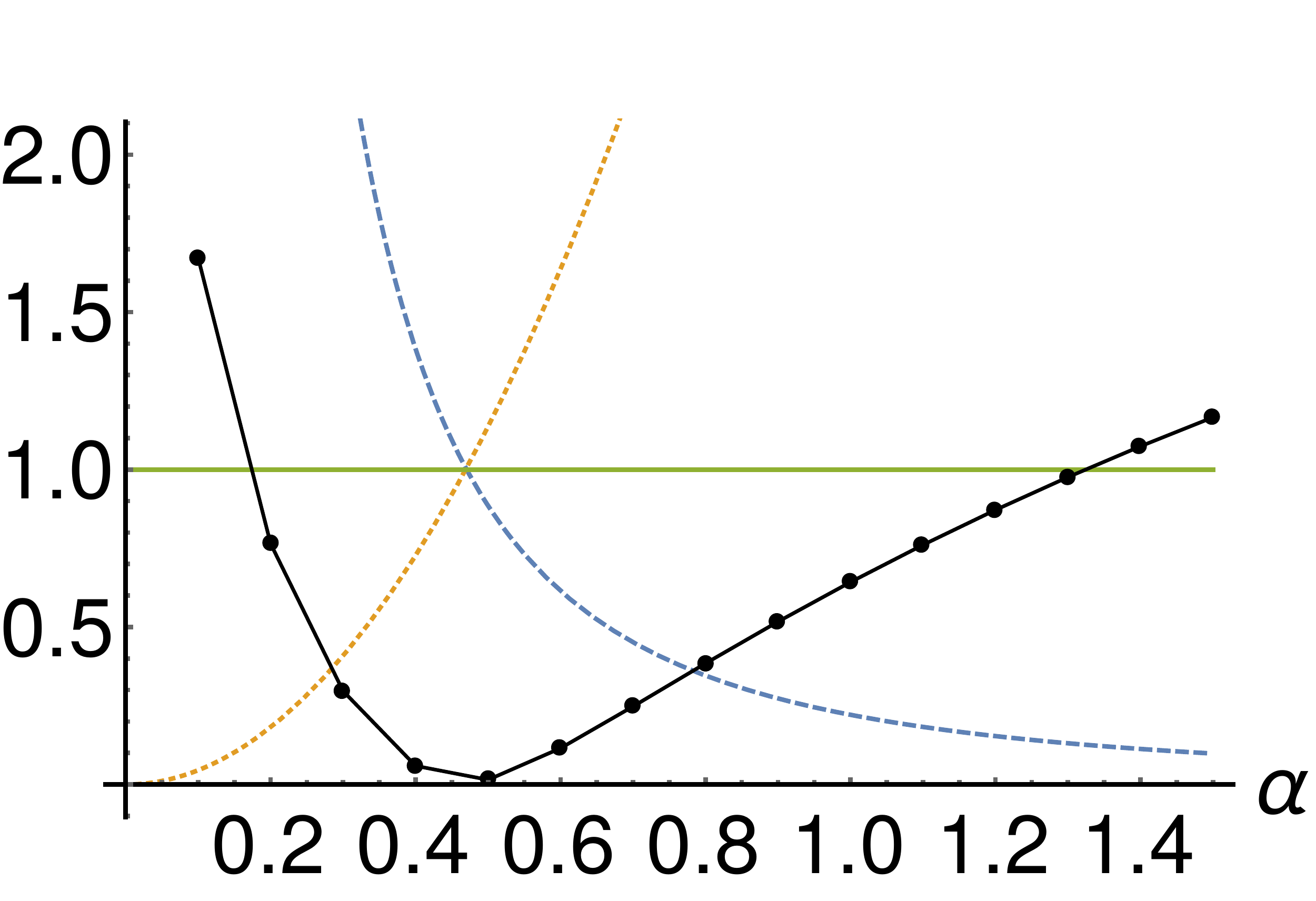}
{\bf (c)}\hskip0.5\columnwidth{\bf (d)}\\
\includegraphics[width=.49\columnwidth]{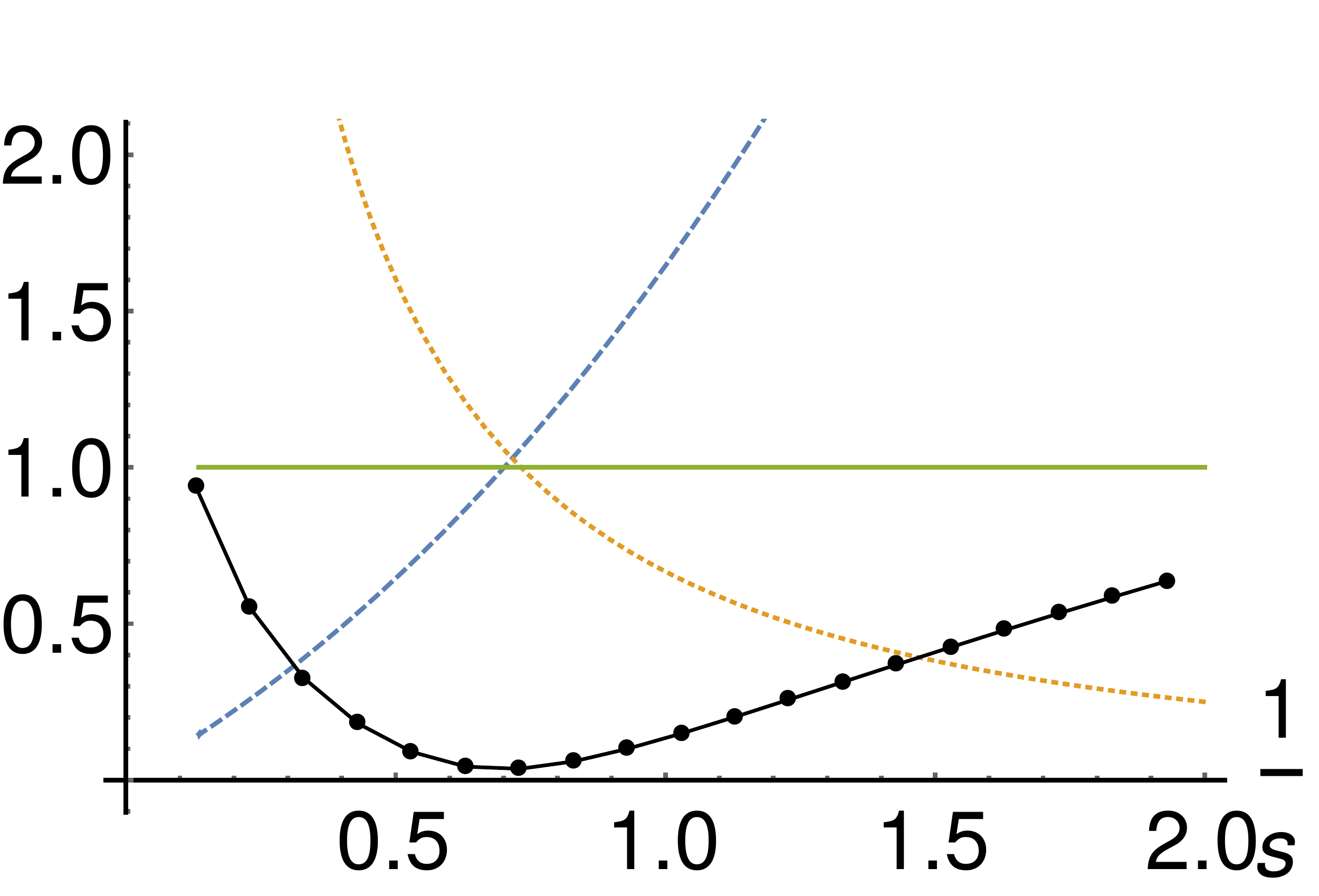}
\includegraphics[width=.49\columnwidth]{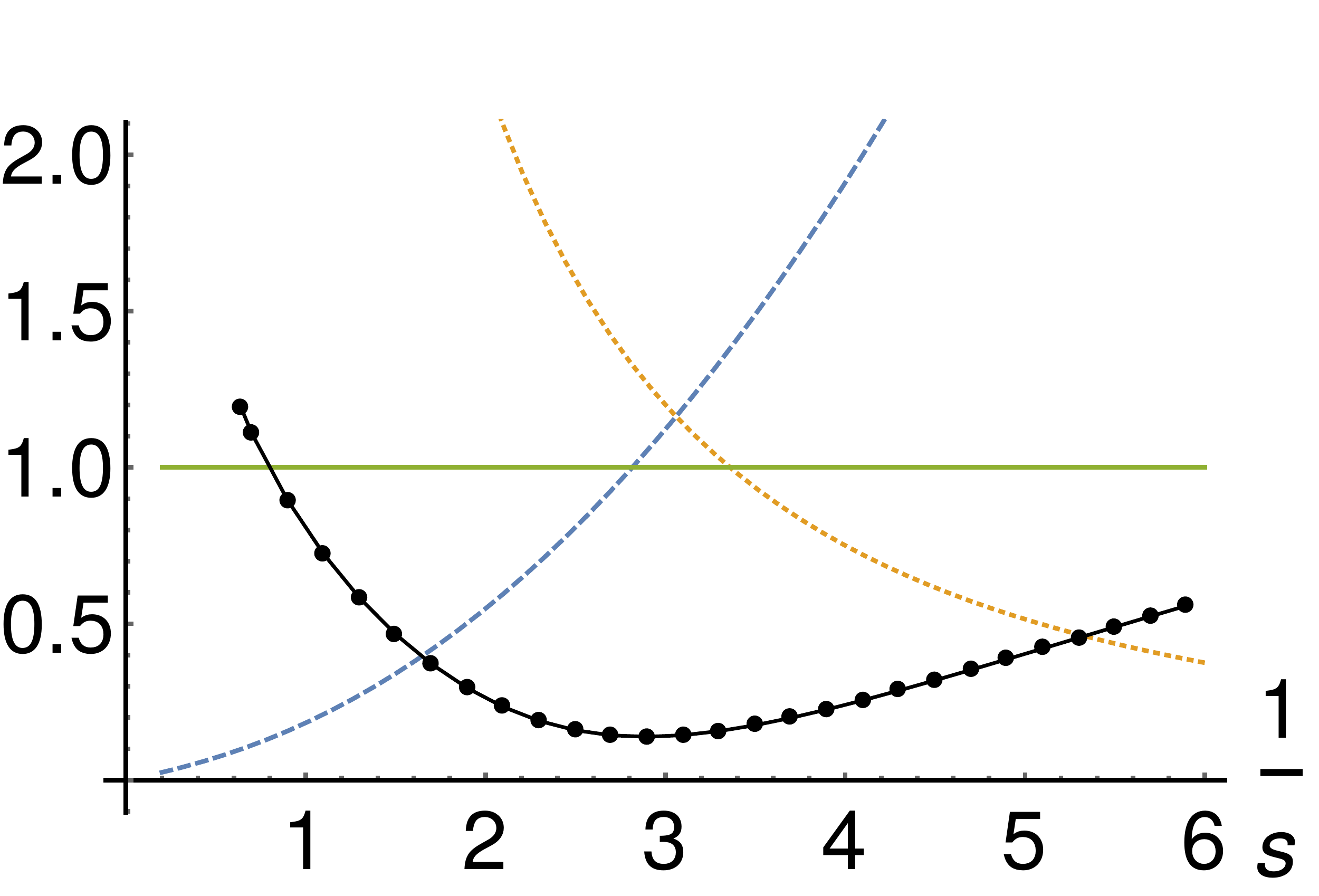}
\caption{(Color online) Entanglement potential and squeezing for
the PT oscillator. Panels {\bf (a)} and {\bf (b)} [{\bf (c)} and {\bf (d)}] show $r_x$ (dashed blue curve),
$r_p$ (dotted orange curves), and the entanglement potential $\ep$ (black dots)against $\alpha$ [$1/s$] for $s=1$ and
$s=5$ [$\alpha=1$ and $\alpha=3$].} \label{fig:PTsqueezingvsEP}
\end{figure}

As for the entanglement potential, this turns out to depend on both $\alpha$ and $s$. Plots similar to those valid for the MHO and Morse potential are presented in Fig.~\ref{fig:PTsqueezingvsEP} (without the $W$-nonclassicality $\nu$).

\section{Oscillators with polynomial perturbations}
\label{sec:harmoscpert}
So far we have studied exactly solvable potentials with two parameters
and revealed a common behavior: the nonlinearity and the $W$-nonclassicality 
$\nu$ have the same behavior and depend just on a single effective 
parameter.  On the other hand, the entanglement potential carries a 
dependence on both the parameters and its different behavior may be 
understood in terms of the squeezing of the state.

Now we want to address the case of a generic two-parameter perturbation, 
so we study a physical system composed of a one-dimensional harmonic oscillator 
with perturbations proportional to $x^4$ and $x^6$ respectively. 
%as usual we work with unitary mass and units rescaled in such a way that $m=1$ and $\hbar = 1$. 
The Hamiltonian of this system thus reads
\begin{equation}
\label{eq:H}
H = \frac{1}{2} (p^2 + \omega^2 x^2) + \epsilon_4 x^4 +
\epsilon_6 x^6\,.
\end{equation}
As the model is not exactly solvable, the properties of the system will be studied using perturbation theory. 
We notice that Eq.~\eqref{eq:H} may also serve as an approximation
for any symmetric (even) potential. We do not consider 
odd powers of $x$, as they give rise to known problems in the 
convergence of the perturbative series. A remark is in order: terms proportional to $x$ and to $x^2$ could in
principle be treated in a perturbative way as well. However, they do not give rise to 
truly anharmonic behavior, and will not be considered in this context.

\label{subsec:PHOApproxSol}
\label{sec:pertstates}
In order to get insight into the ground states for these Hamiltonians we use
first-order time-independent perturbation theory~\cite{Sakurai2011}.  
The state takes the form
\begin{equation}
\label{eq:psi}
|\psi\rangle=\sum^3_{n=0}\gamma_{2n}|2n\rangle,
%|\psi \rangle = \gamma_0 |0\rangle + 
%\gamma_2 |2\rangle + \gamma_4 |4\rangle + \gamma_6 |6\rangle,
\end{equation}
where $|k\rangle$ denotes a Fock number state of the harmonic oscillator 
and the coefficients $\gamma_k$ are in given Appendix~\ref{app:states}.

\begin{figure}[t!]\centering
\includegraphics[width=.5\textwidth]{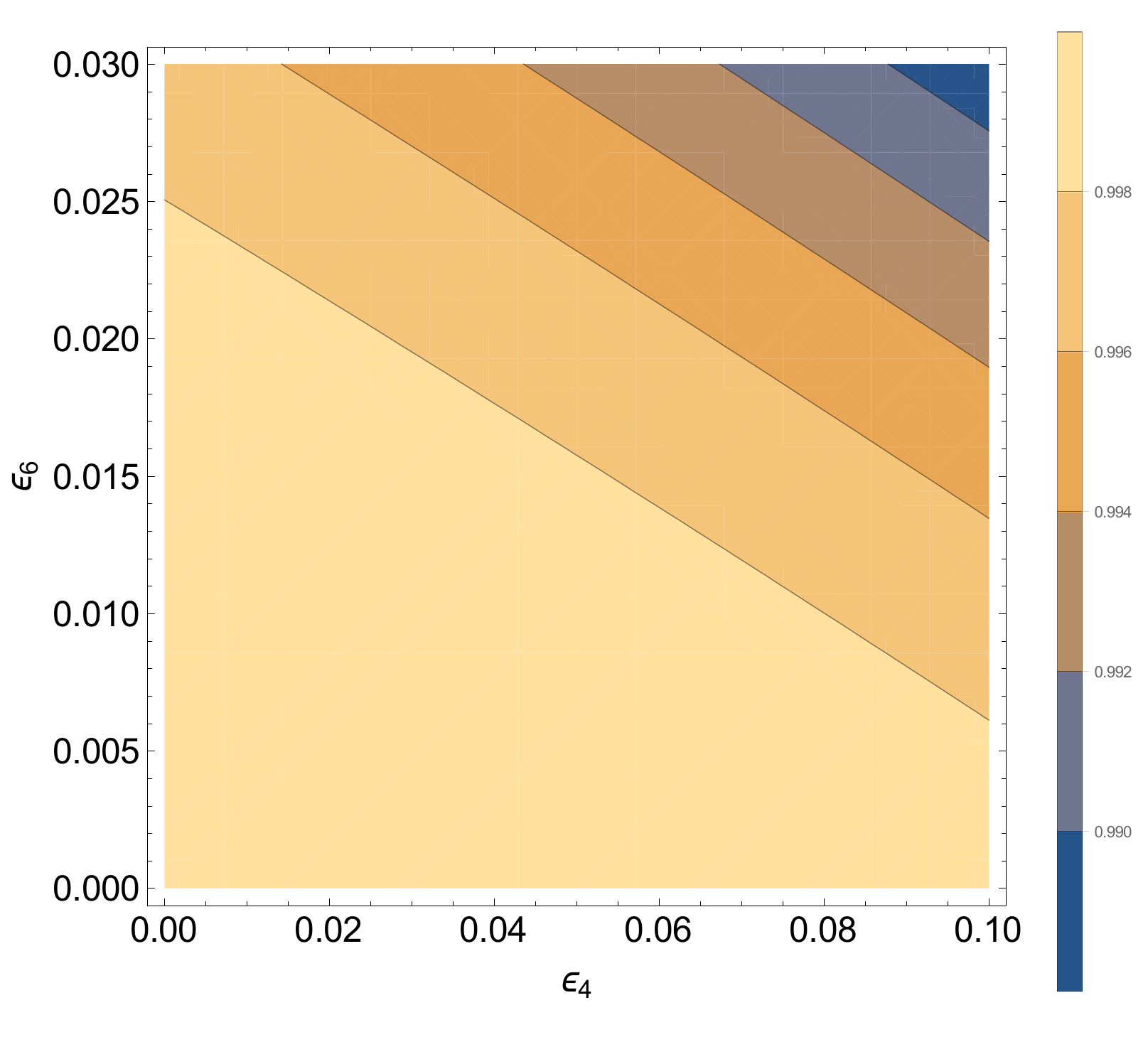}
\caption{(Color online) Contour plot of the overlap between the
perturbative ground states of Eq.~(\ref{eq:psi}) and the numerically calculated one (for $\omega = 1$).} 
\label{fig:OverlapContourPert60}
\end{figure}

First-order perturbation theory returns the ground as a finite
superposition of Fock states, which makes the
Wigner function and the nonlinearity easy to compute. 
In order to assess the validity of the first order approximation,
we compare such ground state to the state obtained by numerically diagonalizing 
the Hamiltonian of the system within a truncated Fock space 
of suitable size. Convergence of the results  of such numerical calculations appear to be ensured by 
using $61$ harmonic levels. The corresponding ground state $|\phi\rangle$ is then compared to $|\psi\rangle$ using the 
state fidelity 
%The matrix elements of the operator $\hat{x}$ are
%\begin{equation}
%x_{ij}=\frac{\sqrt{j} \delta _{i-j+1}+\sqrt{j+1} 
%\delta _{-i+j+1}}{\sqrt{2\omega}},
%\end{equation}
%where $i$ and $j$ go from $0$ to $N$; the 
%operators $\hat{x}^4$ and $\hat{x}^6$ can be easily 
%obtained by standard matrix multiplication. 
%In this basis the matrix representation of the 
%unperturbed Hamiltonian operator $\hat{H}^{(0)}$ is diagonal and its elements are
%\begin{equation}
%H_{jj}^{(0)}=\omega \left( j + \frac{1}{2} \right).
%\end{equation}
%The ground state of the system can thus be calculated by diagonalizing
%the matrix $H^{(0)} + \epsilon_4 x^4 + \epsilon_6 x^6$, which can always
%be done numerically.  The resulting energy eigenstates and eigenvalues
%will be dependent on the dimension $N$ of the Fock space, but they
%should converge for $N$ big enough.  This is indeed the case and we
%choose $N=60$ to obtain a ground state $\phi$ to be confronted with the
%perturbative state $\psi$ using the fidelity 
$|\langle \phi | \psi \rangle |^2$. In Fig.~\ref{fig:OverlapContourPert60} we present a contour plot of
the overlap as a function of both $\epsilon_4$ and $\epsilon_6$.  
For values of $\epsilon_4$ up to 0.1 and $\epsilon_6$ up to 0.03 the fidelity is at least $\approx 0.976$.

\subsection{Nonclassicality and Nonlinearity}
\label{sec:PHOnonlin}
From the perturbed ground state in Eq.~\eqref{eq:psi} we compute the nonlinearity of the perturbing potential. The covariance matrix associated with $|\psi\rangle$ can be thus written as % in terms of the expectation values of creation and destruction operators and
\begin{equation}	
\sigma^{\rm pol}=
\begin{pmatrix}
\frac{(1 + 2 \langle \hat{a}^2 \rangle + 2 \langle \hat{a}^\dag a \rangle - 4 {\langle \hat{a} \rangle}^2 )}{2\omega} & 0 \\
0 & \frac{\omega}{2}(1 + 2 \langle \hat{a}^\dag \hat{a} \rangle - 2 \langle \hat{a}^2 \rangle)
\end{pmatrix},
\end{equation}
% \begin{equation}
% \label{eq:covmatrix_aadag}
% \scalebox{0.6}{\mbox{\ensuremath{\displaystyle
% \begin{pmatrix}
% \frac{1}{2\omega}(1 + \langle \hat{a}^2 \rangle + \langle \left( \hat{a}^\dag\right)^2 \rangle + 2 \langle \hat{a}^\dag \hat{a} \rangle - 2 \langle \hat{a} \rangle \langle \hat{a}^\dag \rangle - {\langle \hat{a} \rangle}^2 -   {\langle \hat{a}^\dag \rangle}^2  ) & \frac{\I}{2} ( \langle \left( \hat{a}^\dag\right)^2  \rangle - \langle \hat{a}^2 \rangle - {\langle \hat{a}^\dag \rangle}^2 + {\langle \hat{a} \rangle}^2 ) \\
% \frac{\I}{2} ( \langle \left( \hat{a}^\dag\right)^2 \rangle - \langle  \hat{a}^2 \rangle - {\langle \hat{a}^\dag \rangle}^2 + {\langle \hat{a} \rangle}^2 ) & \frac{\omega}{2}(1 - \langle  \hat{a}^2 \rangle - \langle  \left( \hat{a}^\dag\right)^2 \rangle \rangle + 2 \langle \hat{a}^\dag \hat{a} \rangle - 2 \langle \hat{a} \rangle \langle \hat{a}^\dag \rangle + {\langle \hat{a} \rangle}^2 +   {\langle \hat{a} 	^\dag \rangle}^2  )
% \end{pmatrix} }}},
% \end{equation}
with $\hat a$ and $\hat a^\dag$ the annihilation and creation operators of the oscillator and 
%these values are easily expressed in terms of the coefficients of \eqref{eq:psi}. 	
\begin{equation}
\label{eq:meanvalues}
\begin{split}
\langle  \hat{a}^\dag \rangle &=  \langle  \hat{a} \rangle = 0, \\
\langle  \hat{a}^\dag a \rangle &= 2 |\gamma_2|^2 + 4 |\gamma_4|^2 + 6  |\gamma_4|^2, \\
\langle  \hat{a}^2  \rangle &= \sqrt{2} \gamma _2 \gamma _0{}^*+2 \sqrt{3} \gamma _4 \gamma _2{}^*+\sqrt{30} \gamma _6 \gamma _4{}^*, \\
\langle  \hat{a}^{\dag2}  \rangle &= \langle  \hat{a}^2  \rangle^* = \sqrt{2} \gamma _0 \gamma _2{}^*+2 \sqrt{3} \gamma _2 \gamma _4{}^*+\sqrt{30} \gamma _4 \gamma _6{}^*.
\end{split}
\end{equation}
%These turn out to be real, since the coefficients $\gamma_i$ of the ground state are real, so we have that $\langle a \rangle = \langle a^\dag \rangle$ and $ \langle a^2 \rangle = \langle (a^\dag)^2 \rangle$. This fact leads to a simplified correlation matrix which is diagonal
%\begin{equation}	
%\begin{pmatrix}
%\frac{1}{2\omega}(1 + 2 \langle \hat{a}^2 \rangle + 2 \langle \hat{a}^\dag a \rangle - 4 {\langle \hat{a} \rangle}^2 ) & 0 \\
%0 & \frac{\omega}{2}(1 + 2 \langle \hat{a}^\dag \hat{a} \rangle - 2 \langle \hat{a}^2 \rangle)
%\end{pmatrix},
%\end{equation}
An explicit calculation shows that the determination of $\sigma^{\rm pol}$, and in turn %so that the determinant is easily computed as a function of the coefficients by inserting \eqref{eq:meanvalues}
%\begin{equation}
%\label{eq:detsigmaexpl}
%\begin{split}
%\det \vec{\sigma} &= \Biggl(  
%4 \gamma _2^4-2 \gamma _0^2 \gamma _2^2+4 \gamma _4^2 \gamma _2^2+24 \gamma _6^2 \gamma _2^2-4 \sqrt{6} \gamma _0 \gamma _4 \gamma _2^2 \\
%&+2 \gamma _2^2-12 \sqrt{10} \gamma _4^2 \gamma _6 \gamma _2-4 \sqrt{15} \gamma _0 \gamma _4 \gamma _6 \gamma _2 \\
%&+16 \gamma _4^4+36 \gamma _6^4+4 \gamma _4^2+18 \gamma _4^2 \gamma _6^2+6 \gamma _6^2+\frac{1}{4}
%\Biggr). 
%\end{split}
%\end{equation}
the nonlinearity $h(\sqrt{\det \vec{\sigma}})$, depends on both the perturbative parameters and on the frequency $\omega$. No single-parameter rescaling can be identified in this case, thus entailing the double-dependence highlighted above, which is passed to the $W$-nonclassicality $\nu$ %we need to integrate the absolute
%value of the Wigner function; details about its form are left in
[cf. Appendix~\ref{app:states}]. %Notice that no common scaling may be found in this 
%case and $\nu$ depends on $\epsilon_4$ and $\epsilon_6$ separately.

%The P-nonclassicality measure $\ep$ can be calculated exactly (up to a
%numerical matrix diagonalization), as our ground state is already a
%linear combination of Fock states, so no approximation is needed.

%\subsubsection{Nonclassicality Versus Nonlinearity}
As our aim is to highlight the role played by the perturbative parameters, in the remainder of our analysis we set $\omega=1$ and %If we deal with the ground states where one of the two parameters is fixed 
%$\eta_{\text{NG}}$, $\ep$ and $\nu$ are all monotonous increasing functions of the single perturbation parameter, as expected. 
%Anyway the full ground state $\psi$ depends on two perturbative parameters, 
%it is then interesting to see how nonclassicality and nonlinearity behave when treated as functions of both parameters. 
generate random pairs of values $(\epsilon_4,\epsilon_6)$ (within the appropriate range of validity of the first-order perturbative approach discussed above) that are then used to compute both the nonclassicality and nonlinearity indicators. 

\begin{figure}[t!]
	\centering
	\includegraphics[width=.5\textwidth]{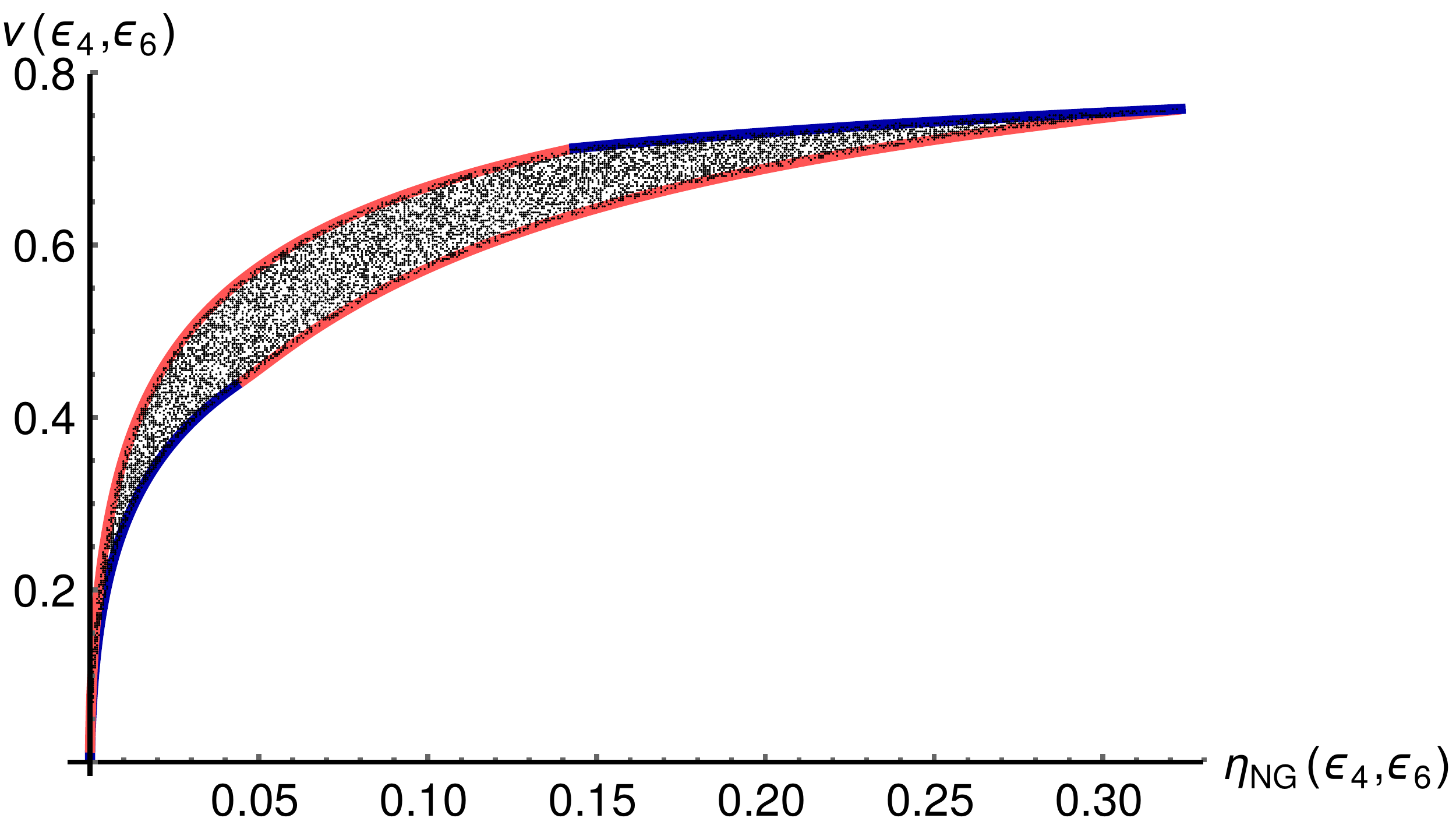}
	\caption{(Color online) Random scatter plot of the $W$-nonclassicality $\nu$ versus the nonlinearity $\eta_{\text{NG}}$ for the perturbed harmonic oscillator when both parameters $\epsilon_4$ and $\epsilon_6$ are are varied in the range given in Fig.~\ref{fig:OverlapContourPert60}; 1000 random points were generated. The dark blue curve below the points represents $\epsilon_6 = 0$, while the one above the points is the curve for $\epsilon_6 = 0.03$. The light red curve below the points is the one for $\epsilon_4 = 0.1$, while the one above the points is for $\epsilon_4 =0$. } 
	\label{fig:PHOdng46}
\end{figure}

\begin{figure}[t!]
	\centering
	\includegraphics[width=.5\textwidth]{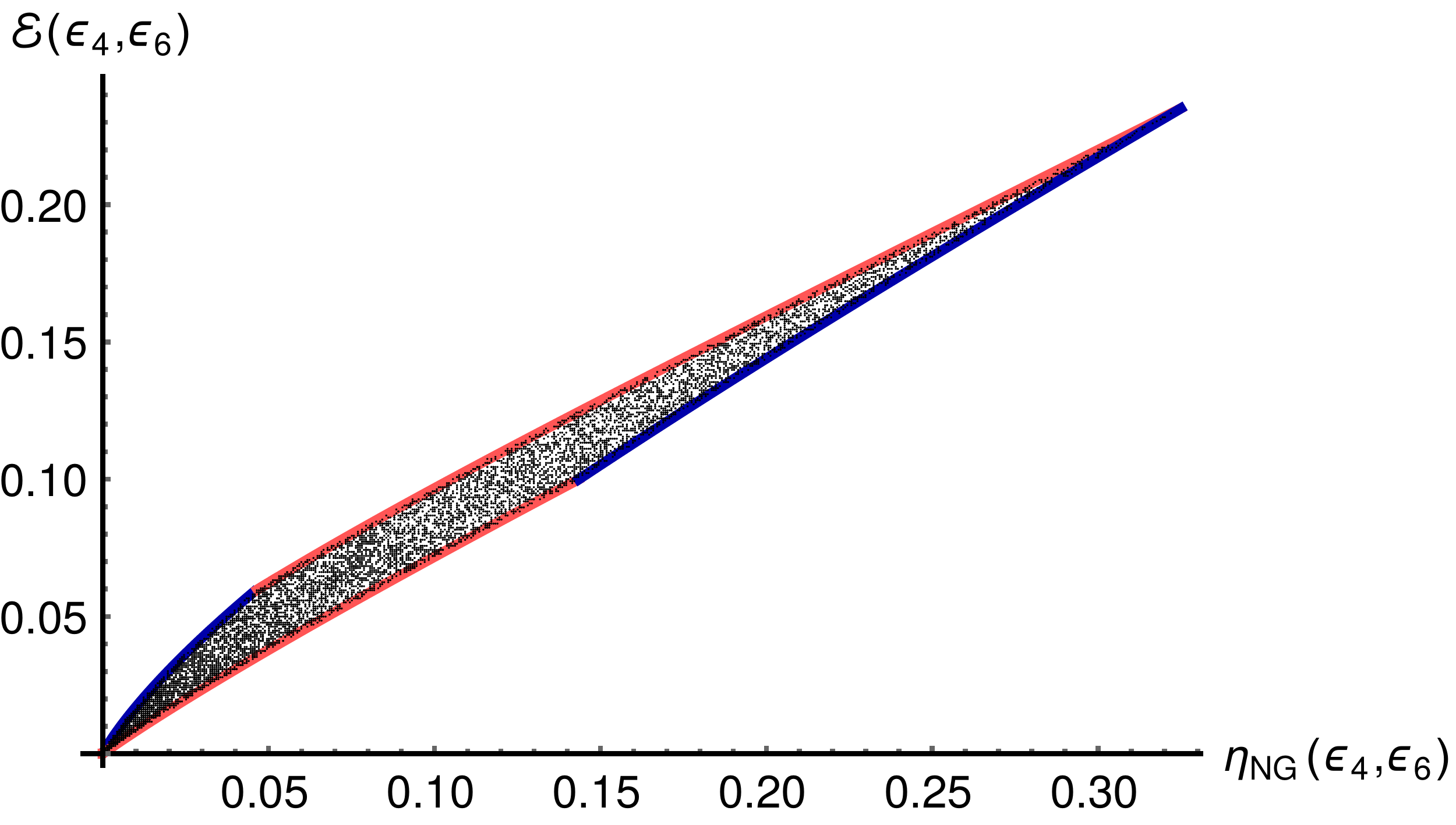}
	\caption{(Color online) Random scatter plot of the entanglement potential (P-nonclassicality) $\ep$ versus the nonlinearity $\eta_{\text{NG}}$ for the perturbed harmonic oscillator when both parameters $\epsilon_4$ and $\epsilon_6$ are varied in the range given in Fig.~\ref{fig:OverlapContourPert60}; 10000 random points were generated. The dark blue curve above the points represents $\epsilon_6 = 0$, while the one below the points is the curve for $\epsilon_6 = 0.03$. The light red curve above the points is the one for $\epsilon_4 = 0.1$, while the one below the points is for $\epsilon_4 =0$.} 
	\label{fig:PHOepng46}
\end{figure}

The results shown in Figs.~\ref{fig:PHOdng46} and~\ref{fig:PHOepng46} showcase a non-monotonic relation between nonlinearity and nonclassicality: the points corresponding to the randomly taken pairs of values for the parameters are distributed within a (narrow) region comprised within four curves, each associated with an extremal value of $\epsilon_{4,6}$.
Nonclassicality and nonlinearity are thus strongly dependent on the
details of the system under consideration and are, strictly speaking, 
non equivalent notions. 
%This means that for a given
%value of nonlinearity the amount of nonclassicality generated is
%different for different values of the parameters. On the other hand, 
On the other hand, the regions in Fig.~\ref{fig:PHOdng46}
and~\ref{fig:PHOepng46} are concentraded enough to suggest 
that the intuitive link between such two features is, in 
fact, correct. Notice also that upon fixing the value of 
one of the parameters (either $\epsilon_4$ or 
$\epsilon_6$) the behaviour of both nonclassicality measures
becomes monotonic with nonlinearity.

Digging into the details of the phenomenological behavior identified by
our analysis, it appears that $W$-nonclassicality
(Fig.~\ref{fig:PHOdng46}) is favoured by the $x^6$-like nonlinearity. On
the other hand, % potential yields a more
%nonclassical ground state than all the other states.
%The situation is different if we consider 
P-nonclassicality, appears to benefit from a $\hat x^4$-type of nonlinear effects: % In Fig.~\ref{fig:PHOepng46} the random points are always delimited by the
%four curves, but this plit has a different structure compared to 
%that of $W$-nonclassicality, and we see that 
in Fig.~\ref{fig:PHOepng46} the roles of the dark blue and light red
curves are inverted with respect to Fig.~\ref{fig:PHOdng46}, showing
that, after choosing the parameters $\epsilon_4$ and $\epsilon_6$ in
such a way that the entropic nonlinearity is fixed, the ground state
obtained with the maximum value of $\epsilon_4$ generates more
entanglement than any other one.

\section{Conclusions}
\label{sec:out}
We have addressed in details the role played by the nonlinearity of 
anharmonic potentials in the generation of nonclassicality in their
ground states. In particular, we have shown that nonlinearity plays
a crucial role in the generation of W-nonclassicality, while
P-nonclassicality may be also obtained by potential inducing just 
squeezing.

Our results support the expectation, put forward in
Ref.~\cite{Teklu2015}, that the nonlinearity of a potential is
quantitatively related to the nonclassicality of its ground state and
thus the former feature may be regarded as a resource to generate the
latter one. The strict validity of such expectation, which appears to be
conceptually quite intuitive, is however strongly linked to the specific
details of the Hamiltonian model being addressed. Anharmonic potentials
that can be reduced to a single-parameter dependence give rise, in fact,
to a monotonic relation between nonlinearity and W-nonclassicality.  
Such a correspondence breaks down for effectively
multi-parameter potentials: set values of nonlinearity bound the
possible degrees of nonclassicality of the ground state of a given
anharmonic potential, albeit without {\it determining} it unambiguously. 

Our investigation opens up a series of questions, all linked to the
effective role that non-harmonic oscillators might have in the quantum
technology arena: it would be interesting, for instance, to investigate
whether the enhanced non-classicality achieved, in general, for a
non-null degree of nonlinearity is accompanied by an equally enhanced
degree of coherence in the ground state of the oscillator. Equally
interesting is the question on the actual use that can be made of the
sought nonlinearity in protocols of practical quantum estimation.

\section*{Acknowledgements}
This work was supported by the UK
EPSRC (EP/M003019/1), the John Templeton Foundation (grant ID
43467), the EU through the Collaborative 
Projects QuProCS (Grant 
Agreement 641277) and TherMiQ (Grant Agreement 618074), and by UniMI through the H2020 Transition 
Grant 15-6-3008000-625.

\begin{widetext}
\appendix
\section{Calculations for the harmonic oscillator with perturbations}
\subsection{Perturbative states}
\label{app:states}
The matrix elements of the two perturbations on the basis of the energy eigenstates of the unperturbed system, which in this case are the Fock states $|n\rangle$, are
\begin{align}
\label{eq:matrixx4}
\langle n | \hat{x}^4 | n \rangle &= \frac{6n^2 +6n +3}{4 \omega^2} \\ 
\langle n | \hat{x}^4 | n+4 \rangle &= \frac{\sqrt{ (n+1) (n+2) (n+3) 
(n+4)}}{4 \omega ^2}, \\ 
\langle n | \hat{x}^4 | n+2 \rangle &= \frac{(4n + 6)\sqrt{n(n-1)}}{4 \omega^2},
\end{align}
for $x^4$ and 
\begin{align}
\label{eq:matrixx6}
\langle n | \hat{x}^6 | n \rangle & 
= \frac{5 \left(4 n^3+6 n^2+8 n+3\right)}{8 \omega ^3} \\ 
\langle n | \hat{x}^6 | n+6 \rangle & =  
\frac{\sqrt{(n+1)(n+2)(n+3)(n+4)(n+5)(n+6)}}{8 \omega ^3} \\
\langle n | \hat{x}^6 | n+4 \rangle  &= \frac{3 (2 n+5) 
\sqrt{(n+1)(n+2)(n+3)(n+4)} }{8 \omega ^3} \\ 
\langle n | \hat{x}^6 | n+2 \rangle &= 
\frac{15 \left(n^2+3 n+3\right) \sqrt{(n+1)(n+2)}}{8 \omega ^3}.
\end{align}
for $x^6$ and all the other elements are zero apart from the symmetrical ones (i.e. $\langle n | \hat{x}^4 | n+k \rangle = \langle n+k | \hat{x}^4 | n \rangle$ and $\langle n | \hat{x}^6 | n+k \rangle = \langle n+k | \hat{x}^6 | n \rangle$). 

The formula for the perturbed ground state is the following
\begin{equation}
\label{eq:pert1ord}
|\psi\rangle = |0\rangle + \epsilon \sum_{k \neq 0} |k\rangle \frac{V_{k0}}{-\omega k},
\end{equation}
where $V_{nk} = \langle n^{(0)} |V | k^{(0)} \rangle$ and $V$ stands for the perturbation $\epsilon_4 x^4 + \epsilon_6 x^6$.
Using this formula and the matrix elements found in the last section we readily find the normalized ground state \eqref{eq:psi},
its coefficients are the following
\begin{equation}
\label{eq:coeffpsi}
\begin{split}
\gamma_0 &= \frac{1}{C} \qquad \gamma_2 = -\frac{\gamma_0}{\sqrt 2}  \left( \frac{45 \epsilon _6}{4 \omega ^3}+\frac{3 \epsilon _4}{ \omega ^2} \right) \\
\gamma_4 &= -\gamma_0\sqrt{\frac{3}{2}} \left( \frac{15  \epsilon _6}{2 \omega ^3}+\frac{\epsilon _4}{\omega ^2}  \right)  \qquad 
\gamma_6 = -{ \sqrt{5} }\gamma_0 \epsilon _6,
\end{split}
\end{equation}
where the normalization constant $C$ is 
\begin{equation}
\label{eq:C}
C = \frac{\sqrt{\omega ^2 \left(96 \omega ^6+117 \epsilon _4^2\right)+945 \omega  \epsilon _4 \epsilon _6+2055 \epsilon _6^2}}{4 \sqrt{6} \omega ^4}.
\end{equation}

%\begin{widetext}
\subsection{Wigner function}
It is convenient to express the Wigner function as $W(\alpha) = \frac{2}{\pi} \Tr\left[ \rho \hat{D}(2 \alpha) (-1)^{\hat{a}^{\dag}\hat{a} } \right]$~\cite{Cahill1969a,Barnett1997}
%\begin{equation}
%\label{eq:wigcheuso}
%W(\alpha) = \frac{2}{\pi} \Tr\left[ \rho \hat{D}(2 \alpha) (-1)^{\hat{a}^{\dag}\hat{a} } \right],
%\end{equation}
where $\hat{D}$ is the displacement operator $\hat{D}\left( \xi \right) = \exp\left( \xi \hat{a}^{\dag} - \xi^{*} \hat{a} \right)$. The expectation values of $\hat{D}$ on Fock states are given by
\begin{equation}
\langle n' | D(z) | n \rangle =
  \begin{cases}
    \sqrt{\frac{n!}{n'!}} e^{-\frac{|z|^2}{2}}(-z)^{n'-n}L^{(n'-n)}_{n'}(|z|^2) & \text{if} \quad n' > n \\
    \sqrt{\frac{n'!}{n!}} e^{-\frac{|z|^2}{2}}(z^*)^{n-n'}L^{(n-n')}_{n}(|z|^2) & \text{if} \quad n > n'
  \end{cases}
\end{equation}

where $L_n^{(\alpha)} (x)$ are the associated Laguerre polynomials.
The Wigner function then becomes
\begin{equation}
\label{eq:wigfun}
\begin{aligned}
W(z) &= \frac{2}{\pi} e^{-2|z|^2} \bigl[ {\gamma_0}^2 L_0(4|z|^2) + {\gamma_2}^2 L_2(4|z|^2) 
+ {\gamma_4}^2 L_4(4|z|^2)\\
& +	 {\gamma_6}^2 L_6(4|z|^2) + 4\sqrt{2} \gamma_0 \gamma_2\operatorname{Re}(z^2) L_2^2(4|z|^2)  \\ 
&+\frac{16}{\sqrt{3}} \gamma_0 \gamma_4 \operatorname{Re}(z^4) L_4^4(4|z|^2) + \frac{32}{3\sqrt{5}} \gamma_0 \gamma_6 \operatorname{Re}(z^6) L_6^6(4|z|^2)\\
& + \frac{4}{\sqrt{3}} \gamma_2 \gamma_4 \operatorname{Re}(z^2) L_4^2(4|z|^2) + \frac{16}{3\sqrt{10}} \gamma_2 \gamma_6 \operatorname{Re}(z^4) L_6^4(4|z|^2)\\
& + \frac{8}{\sqrt{30}} \gamma_4 \gamma_6 \operatorname{Re}(z^2) L_6^2(4|z|^2) \bigr],
\end{aligned}
\end{equation}
where the coefficients are given by \eqref{eq:coeffpsi}.
\end{widetext}

%\bibliography{MasterThesis2014}

\begin{thebibliography}{99}
\bibitem{NC10} A.~Nielsen and I.L.~Chuang, \textit{Quantum computation and quantum information} (Cambridge university press, Cambridge, 2010).
\bibitem{MW95} L.~Mandel and E.~Wolf, \textit{Optical coherence and quantum optics} (Cambridge university press, Cambridge, 1995).
\bibitem{L14} For a recent review see A.I.~Lvovsky, \textit{Squeezed Light} in \textit{Photonics Volume 1: Fundamentals of Photonics and Physics} (Wiley, West Sussex, United Kingdom, 2015, pp 121-164).
\bibitem{L03} D.~Leibfried, \textit{et al.}, Rev. Mod. Phys. \textbf{75}, 281 (2003).
\bibitem{AKM14} M.~Aspelmeyer, T.J.~Kippenberg, and F.~Marquardt, Rev. Mod. Phys. \textbf{86}, 1391 (2014).
\bibitem{LSA13} M. Lewenstein, A. Sanpera, and V. Ahufinger, \textit{Ultracold Atoms in Optical Lattices: Simulating quantum many-body systems}, Oxford U. Press (2013).
\bibitem{R+14} B.~Rogers \textit{et al.}, Quantum Measurements and Quantum Metrology \textbf{2}, 11 (2014).
\bibitem{H+11} J. P. Home, D. Hanneke, J. D. Jost, D. Leibfried, and D.J. Wineland, New J. Phys. \textbf{13}, 073026 (2011).
\bibitem{S+09} J.C. Sankey et al., Nature Phys. 6, 707 (2009).
\bibitem{PEANO} V. Peano and M. Thorwart, New J. Phys. \textbf{8}, 21 (2006).
\bibitem{KOLKIRAN} A. Kolkiran and G. S. Agarwal, arXiv:0608621v2.
\bibitem{ANDERSSON} C. Joshi, M. Jonson, E. Andersson, and P. \"{O}hberg, J. Phys. B: At. Mol. Opt. Phys. \textbf{44}, 245503 (2011).
\bibitem{ONG} F. R. Ong, M. Boissonneault, F. Mallet, A. Palacios-Laloy, A. Dewes, A. C. Doherty, A. Blais, P. Bertet, D. Vion, and D. Esteve, Phys. Rev. Lett. \textbf{106}, 167002 (2011).
\bibitem{DIVINCENZO} D. P. DiVincenzo and J. A. Smolin, New J. Phys. \textbf{14}, 013051 (2012).
\bibitem{RIPS} S. Rips and M. J. Hartmann, Phys. Rev. Lett. \textbf{110}, 120503 (2013).
\bibitem{Vacanti2013} G. Vacanti, M. Paternostro, G. M. Palma, M. S. Kim, and V. Vedral, Phys. Rev. A {\bf 88}, 013851 (2013).
\bibitem{MFB14} V.~Montenegro, A.~Ferraro, and S.~Bose, Phys. Rev. A \textbf{90}, 013829 (2014).
\bibitem{H87} M. Hillery, Phys. Rev. A \textbf{35}, 725 (1987).
\bibitem{Lee1991} C. T. Lee, Phys. Rev. A {\bf 44}, 2775(R) (1991).
\bibitem{RV02} T.~Richter and W.~Vogel, Phys. Rev.
Lett. \textbf{89}, 283601 (2002).
\bibitem{Kenfack2004} A. Kenfack and K. $\dot{\rm Z}$yczkowski, J. Opt. B {\bf 6}, 396 (2004).
\bibitem{Asboth2005} J. Asb\'oth, J. Calsamiglia, and H. Ritsch, Phys. Rev. Lett. {\bf 94}, 173602 (2005).
\bibitem{Mari2011} A. Mari, K. Kieling, B. M. Nielsen, E. S. Polzik, and J. Eisert, Phys. Rev. Lett. {\bf 106}, 010403 (2011).
\bibitem{FP12} A.~Ferraro and M.G.A.~ Paris , Phys. Rev. Lett. {\bf 108}, 260403 (2012).
\bibitem{Teklu2015} B. Teklu, A. Ferraro, M. Paternostro, and M. G. A. Paris, (2015), EPJ Quantum Technol. {\bf 2}, 16 (2015).
\bibitem{Paris2014} M. G. A. Paris, M. G. Genoni, N. Shammah, and B. Teklu, Phys. Rev. A {\bf 90}, 012104 (2014).
\bibitem{Genoni2010} M. G. Genoni and M. G. A. Paris, Phys. Rev. A {\bf 82}, 052341 (2010).
\bibitem{Li2010} J. Li, G. Li, J.-M. Wang, S.-Y. Zhu,and T.-C. Zhang, J. Phys. B {\bf 43}, 085504 (2010).
\bibitem{Miranowicz2015} A. Miranowicz, K. Bartkiewicz, A. Pathak, J. J. Perina, Y.-N. Chen, and F. Nori, (2015), arXiv:1502.04523v1.
\bibitem{Glauber1963} R. J. Glauber, Phys. Rev. {\bf 130} 2529. {\bf 49} (1963).
\bibitem{Titulaer1965} U. M. Titulaer and R. J. Glauber, Phys. Rev. {\bf 140}, B676 (1965).
\bibitem{Mandel1986} L. Mandel, Phys. Scr. {\bf 1986}, 34 (1986).
\bibitem{Vogel2006} W. Vogel and D.-G. Welsch, Quantum Optics, 3rd, Revised and Extended Edition (Wiley-Vch, Verlang Berlin GmbH, 2006) p. 520.
\bibitem{Johansen2004} L. M. Johansen, Phys. Lett. A {\bf 329}, 184 (2004).
\bibitem{Spekkens2007} R. W. Spekkens, Phys. Rev. Lett. {\bf 101}, 020401 (2008).
\bibitem{Kiesel2013} T. Kiesel, Phys. Rev. A {\bf 87}, 062114 (2013).
\bibitem{Lutkenhaus1995} N. L\"utkenhaus, and S. M. Barnett, Phys. Rev. A {\bf 51}, 3340 (1995).
\bibitem{Aharonov1966} Y. Aharonov, D. Falkoff, E. Lerner, and H. Pendleton, Ann. Phys. USA {\bf 39}, 498 (1966).
\bibitem{Kim2002} M. S. Kim, W. Son, V. Bu\v{z}ek, and P. L. Knight, Phys. Rev. A {\bf 65}, 032323 (2002).
\bibitem{Xiang-bin2002} W. Xiang-bin, Phys. Rev. A {\bf 66}, 024303 (2002).
%\bibitem{Usenko2010} V. Usenko and M. G. A. Paris, Phys. Lett. A {\bf 374}, 1342 (2010).
\bibitem{WEP:03} M.M.~Wolf, J.~Eisert, and M.B.~Plenio, Phys.~Rev.~Lett.
\textbf{90}, 047904 (2003).
\bibitem{oli09} S. Olivares and M. G. A. Paris, 
Phys. Rev. A {\bf 80}, 032329 (2009).
\bibitem{oli11} S. Olivares and M. G. A. Paris, Phys. Rev. 
Lett. {\bf 107}, 170505 (2011).
\bibitem{JLC:13} Z.~Jiang, M.D.~Lang, and C.M.~Caves, Phys. Rev. A
\textbf{88}, 044301 (2013).
\bibitem{VS:14} W.~Vogel and J.~Sperling, Phys. Rev. A 
\textbf{89}, 052302 (2014).
\bibitem{B+15} M.~Brunelli \textit{et al.}, Phys. Rev. A \textbf{91}, 062315 (2015).
\bibitem{Mari2012} A. Mari and J. Eisert, Phys. Rev. Lett. {\bf 109}, 230503 (2012).
\bibitem{Veitch2013} V. Veitch, N. Wiebe, C. Ferrie, and J. Emerson, New J. Phys. {\bf 15}, 013037 (2013).
\bibitem{Hudson1974} R. L. Hudson, Reports Math. Phys. {\bf 6}, 294 (1974).
\bibitem{Genoni2008} M. G. Genoni, M. G. A. Paris, and K. Banaszek, Phys. Rev. A {\bf 78}, 060303 (2008).
\bibitem{Ferraro2005} A. Ferraro, S. Olivares, and M. G. A. Paris, {\it Gaussian states in continuous variable quantum information} (Bibliopolis, Napoli, 2005).
\bibitem{Bund2000} G. W. Bund, and M. C. Tijero, Phys. Rev. A {\bf 61}, 1 (2000).
\bibitem{Morse1929} P. M. Morse, Phys. Rev. {\bf 34}, 57 (1929).
\bibitem{Frank2000} A. Frank, A. Rivera, and K. B. Wolf, Phys. Rev. A {\bf 61}, 054102 (2000).
\bibitem{Hahn2005} T. Hahn, Comput. Phys. Commun. {\bf 176}, 712 (2007).
\bibitem{Sakurai2011} J. J. Sakurai and J. Napolitano, {\it Modern quantum mechanics}, 2nd ed. (Addison-Wesley, Boston, 2011).
\bibitem{Cahill1969a} K. E. Cahill and R. J. Glauber, Phys. Rev. {\bf 177}, 1882 (1969).
\bibitem{Barnett1997} S. M. Barnett and P. M. Radmore, {\it Methods in theoretical quantum optics} (Oxford University Press, Oxford, New York, 1997).
\end{thebibliography}
\end{document}